\documentclass[11pt,journal,epsfig,onecolumn,peerreviewca,draftclsnofoot]{IEEEtran}

\usepackage[dvips]{graphicx}
\usepackage{graphicx}
\usepackage{amssymb}
\usepackage{cite}
\usepackage{subfigure}
\usepackage{amsmath}

\begin{document}

\title{Super-Resolution Compressed Sensing: A Generalized Iterative Reweighted $\ell_2$ Approach}


\author{Jun Fang, Huiping Duan, Jing Li, Hongbin Li,~\IEEEmembership{Senior
Member,~IEEE}, and Rick S. Blumn,~\IEEEmembership{Fellow,~IEEE}
\thanks{Jun Fang, and Jing Li are with the National Key Laboratory on Communications,
University of Electronic Science and Technology of China, Chengdu
611731, China, Email: JunFang@uestc.edu.cn}
\thanks{Huiping Duan is with the School of Electronic Engineering,
University of Electronic Science and Technology of China, Chengdu
611731, China, Email: huipingduan@uestc.edu.cn}
\thanks{Hongbin Li is
with the Department of Electrical and Computer Engineering,
Stevens Institute of Technology, Hoboken, NJ 07030, USA, E-mail:
Hongbin.Li@stevens.edu}
\thanks{Rick S. Blum is with the Department of Electrical and Computer Engineering, Lehigh University,
Bethlehem, PA 18015, USA, E-mail: rblum@lehigh.edu}
\thanks{This work was supported in part by the National Science
Foundation of China under Grant 61172114, and the National Science
Foundation under Grants ECCS-1408182 and ECCS-1405579. }}

\maketitle

\begin{abstract}
Conventional compressed sensing theory assumes signals have sparse
representations in a known, finite dictionary. Nevertheless, in
many practical applications such as direction-of-arrival (DOA)
estimation and line spectral estimation, the sparsifying
dictionary is usually characterized by a set of unknown parameters
in a continuous domain. To apply the conventional compressed
sensing technique to such applications, the continuous parameter
space has to be discretized to a finite set of grid points, based
on which a ``presumed dictionary'' is constructed for sparse
signal recovery. Discretization, however, inevitably incurs errors
since the true parameters do not necessarily lie on the
discretized grid. This error, also referred to as grid mismatch,
may lead to deteriorated recovery performance or even recovery
failure. To address this issue, in this paper, we propose a
generalized iterative reweighted $\ell_2$ method which jointly
estimates the sparse signals and the unknown parameters associated
with the true dictionary. The proposed algorithm is developed by
iteratively decreasing a surrogate function majorizing a given
objective function, resulting in a gradual and interweaved
iterative process to refine the unknown parameters and the sparse
signal. A simple yet effective scheme is developed for adaptively
updating the regularization parameter that controls the tradeoff
between the sparsity of the solution and the data fitting error.
Extension of the proposed algorithm to the multiple measurement
vector scenario is also considered. Numerical results show that
the proposed algorithm achieves a super-resolution accuracy and
presents superiority over other existing methods.
\end{abstract}




\section{Introduction}
Compressed sensing finds a variety of applications in practice as
many natural signals admit a sparse or an approximate sparse
representation in a certain basis. Nevertheless, accurate
reconstruction of a sparse signal relies on the knowledge of the
sparsifying dictionary, while in many applications, it is often
impractical to pre-specify a dictionary that can sparsely
represent the signal. For example, for the line spectral
estimation problem, using a preset discrete Fourier transform
(DFT) matrix suffers from considerable performance degradation
because the true frequency components may not lie on the
pre-specified frequency grid \cite{TangBhaskar12,CandesGranda12}.
The same is true for direction-of-arrival (DOA) estimation and
source localization in sensor networks, where the true directions
or locations of sources may not be aligned on the presumed grid
\cite{YangXie13}. Overall, in these applications, the sparsifying
dictionary is characterized by a set of unknown parameters in a
continuous domain. In order to apply compressed sensing to such
applications, the continuous parameter space has to be discretized
to a finite set of grid points, based on which a presumed
dictionary is constructed for sparse signal recovery.
Discretization, however, inevitably incurs errors since the true
parameters do not necessarily lie on the discretized grid. This
error, also referred to as the grid mismatch, leads to
deteriorated performance or even failure in recovering the sparse
signal. Finer grids can certainly be used to reduce grid mismatch
and improve the reconstruction accuracy. Nevertheless, recovery
algorithms may become numerically instable and computationally
prohibitive when very fine discretized grids are employed.

The grid mismatch problem has attracted a lot of attention over
the past few years. Specifically, in \cite{ChiScharf11}, the
problem was addressed in a general framework of ``basis mismatch''
where the mismatch is modeled as a perturbation (caused by grid
discretization, calibration errors or other factors) between the
presumed and the actual dictionaries, and the impact of the basis
mismatch on the reconstruction error was analyzed. In
\cite{YangXie13,HuZhou13}, to deal with grid mismatch, the true
dictionary is approximated as a summation of a presumed dictionary
and a structured parameterized matrix via the Taylor expansion.
The recovery performance of this method, however, depends on the
accuracy of the Taylor expansion in approximating the true
dictionary. The grid mismatch problem was also examined in
\cite{FannjiangLiao12,DuarteBaraniuk13}, where a highly coherent
dictionary (very fine grids) is used to mitigate the
discretization error, and a class of greedy algorithms which use
the technique of band exclusion (coherence-inhibiting) were
proposed for sparse signal recovery. Besides these efforts,
another line of work \cite{HuShi12,CandesGranda12,TangBhaskar12}
studied the problem of grid mismatch in a more fundamental way:
they circumvent the discretization issue by working directly on
the continuous parameter space, leading to the so-called
super-resolution technique. In
\cite{CandesGranda12,TangBhaskar12}, an atomic norm-minimization
(also referred to as the total variation norm-minimization)
approach was proposed to handle the infinite dictionary with
continuous atoms. It was shown that given that the frequency
components are sufficiently separated, the frequency components of
a mixture of complex sinusoids can be super-resolved with infinite
precision from coarse-scale information only. Nevertheless,
finding a solution to the atomic norm problem is quite
challenging. Although the atomic norm problem can be cast into a
convex semidefinite program optimization for the complex sinusoid
mixture problem, it still remains unclear how this reformulation
generalizes to other scenarios. In \cite{HuShi12}, by treating the
sparse signal as hidden variables, a Bayesian approach was
proposed to iteratively refine the dictionary, and is shown able
to achieve super-resolution accuracy.







In this paper, we propose a generalized iterative reweighted
$\ell_2$ method for joint dictionary parameter learning and sparse
signal recovery. The proposed method is developed by iteratively
decreasing a surrogate function that majorizes the original
objective function. Note that the use of the iterative reweighted
scheme for sparse signal recovery is not new and has achieved
great success over past few years (e.g.
\cite{GorodnitskyRao97,RaoDelgado99,CandesWakin08,ChartrandYin08,WipfNagaranjan10}).
Nevertheless, previous works concern only recovery of the sparse
signal. The current work, instead, generalizes the iterative
reweighted scheme for joint dictionary parameter learning and
sparse signal recovery. Moreover, previous iterative reweighted
algorithms usually involve iterative minimization of a surrogate
function majorizing a given objective function, while our proposed
method only requires iteratively decreasing a surrogate function.
We will show that through iteratively decreasing (not necessarily
minimizing) the surrogate function, the iterative process yields a
non-increasing objective function value as well, and is guaranteed
to converge to a stationary point of the objective function. This
generalization extends the applicability of the iterative
reweighted scheme since finding a simple and convex surrogate
function which admits an analytical solution could be difficult
for many complex problems. In addition, iteratively decreasing the
surrogate function results in an interweaved and gradual
refinement of the signal and the unknown parameters, which enables
the algorithm to produce more focal and reliable estimates as the
optimization progresses. The current work is an extension of our
previous work \cite{FangLi14} to more general scenarios involving
noisy and/or multiple measurement vectors. As shown in this paper,
this extension is technically non-trivial and also brings in
substantial reduction in computational complexity.



The rest of the paper is organized as follows. In Section
\ref{sec:formulation}, the line spectral estimation problem is
formulated as a joint sparse representation and dictionary
parameter estimation problem. A generalized iterative reweighted
$\ell_2$ algorithm is developed in Section \ref{sec:algorithm}.
The choice of the regularization parameter controlling the
tradeoff between sparsity and data fitting is discussed in Section
\ref{sec:lambda-choice}, where a simple and effective update rule
for the regularization parameter is proposed. Extension of the
proposed algorithm to the multiple measurement vector scenario is
studied in Section \ref{sec:mmv}. In Section \ref{sec:analysis},
we provide a heuristic but enlightening analysis on the exact
reconstruction condition of the considered problem for the
noiseless case. Simulation results are provided in Section
\ref{sec:simulation}, followed by concluding remarks in Section
\ref{sec:conclusion}.

\section{Problem Formulation} \label{sec:formulation}
In many practical applications such as direction-of-arrival (DOA)
estimation and line spectral estimation, the sparsifying
dictionary is usually characterized by a set of unknown parameters
in a continuous domain. For example, consider the line spectral
estimation problem where the observed signal is a summation of a
number of complex sinusoids:
\begin{align}
y_m=\sum_{k=1}^K \alpha_k e^{-j\omega_k m}+w_m \qquad m=1,\ldots,
M \label{data-model}
\end{align}
where $\omega_k\in [0, 2\pi)$ and $\alpha_k$ denote the frequency
and the complex amplitude of the $k$-th component, respectively,
and $w_m$ represents the observation noise. Define
$\boldsymbol{a}(\omega)\triangleq[e^{-j\omega}\phantom{0}e^{-j2\omega}\phantom{0}\ldots\phantom{0}e^{-jM\omega}
]^T$, the model (\ref{data-model}) can be rewritten in a
vector-matrix form as
\begin{align}
\boldsymbol{y}=\boldsymbol{A}(\boldsymbol{\omega})\boldsymbol{\alpha}+\boldsymbol{w}
\label{data-model-vector}
\end{align}
where $\boldsymbol{y}\triangleq
[y_1\phantom{0}\ldots\phantom{0}y_M]^T$,
$\boldsymbol{\alpha}\triangleq
[\alpha_1\phantom{0}\ldots\phantom{0}\alpha_K]^T$, and
$\boldsymbol{A}(\boldsymbol{\omega})\triangleq
[\boldsymbol{a}(\omega_1)\phantom{0}\ldots\phantom{0}\boldsymbol{a}(\omega_K)]$.
Note that in some applications, to facilitate data acquisition and
subsequent processing, we may wish to estimate $\{\omega_k\}$ and
$\{\alpha_k\}$ from a subset of measurements randomly extracted
from $\{y_m\}_{m=1}^M$. This random sampling operation amounts to
retaining the corresponding rows of
$\boldsymbol{A}(\boldsymbol{\omega})$ and removing the rest rows
from the dictionary. This modification, however, makes no
difference to our algorithm development.


We see that the dictionary $\boldsymbol{A}(\boldsymbol{\omega})$
is characterized by a number of unknown parameters $\{\omega_k\}$
which need to be estimated along with the unknown complex
amplitudes $\{\alpha_k\}$. To deal with this problem, conventional
compressed sensing techniques discretize the continuous parameter
space into a finite set of grid points, assuming that the unknown
frequency components $\{\omega_k\}$ lie on the discretized grid.
Estimating $\{\omega_k\}$ and $\{\alpha_k\}$ can then be
formulated as a sparse signal recovery problem
$\boldsymbol{y}=\boldsymbol{Az}+\boldsymbol{w}$, where
$\boldsymbol{A}\in\mathbb{C}^{M\times N}$ ($M\ll N$) is an
overcomplete dictionary constructed based on the discretized grid
points. Discretization, however, inevitably incurs errors since
the true parameters do not necessarily lie on the discretized
grid. This error, also referred to as the grid mismatch, leads to
deteriorated performance or even failure in recovering the sparse
signal.

To circumvent this issue, we treat the overcomplete dictionary as
an unknown parameterized matrix
$\boldsymbol{A}(\boldsymbol{\theta})\triangleq
[\boldsymbol{a}(\theta_1)\phantom{0}\ldots\phantom{0}\boldsymbol{a}(\theta_N)]$,
with each atom $\boldsymbol{a}(\theta_n)$ determined by an unknown
frequency parameter $\theta_n$. Estimating $\{\omega_k\}$ and
$\{\alpha_k\}$ can still be formulated as a sparse signal recovery
problem. Nevertheless, in this framework, the frequency parameters
$\boldsymbol{\theta}\triangleq \{\theta_n\}_{n=1}^N$ need to be
optimized along with the sparse signal such that the parametric
dictionary will approach the true sparsifying dictionary.
Specifically, the problem can be presented as follows: we search
for a set of unknown parameters $\{\theta_n\}_{n=1}^N$ with which
the observed signal $\boldsymbol{y}$ can be represented by as few
atoms as possible with a specified error tolerance. Such a problem
can be readily formulated as
\begin{align}
\min_{\boldsymbol{z},\boldsymbol{\theta}}\quad &
\|\boldsymbol{z}\|_0 \nonumber\\
\text{s.t.} \quad
&\|\boldsymbol{y}-\boldsymbol{A}(\boldsymbol{\theta})\boldsymbol{z}\|_2\leq\varepsilon
\label{opt-1}
\end{align}
where $\|\boldsymbol{z}\|_0$ stands for the number of the nonzero
components of $\boldsymbol{z}$, and $\varepsilon$ is an error
tolerance parameter related to noise statistics. The optimization
(\ref{opt-1}), however, is an NP-hard problem. Thus, alternative
sparsity-promoting functionals which are more computationally
efficient in finding the sparse solution are desirable. In this
paper, we consider the use of the log-sum sparsity-encouraging
functional for sparse signal recovery. Log-sum penalty function
has been extensively used for sparse signal recovery, e.g.
\cite{OlshausenField96,ChartrandYin08}. It was proved
theoretically \cite{ShenFang13} and shown in a series of
experiments \cite{CandesWakin08} that log-sum based methods
present uniform superiority over the conventional $\ell_1$-type
methods. Replacing the $\ell_0$-norm in (\ref{opt-1}) with the
log-sum functional leads to
\begin{align}
\min_{\boldsymbol{z},\boldsymbol{\theta}}\quad &
L(\boldsymbol{z})=\sum_{n=1}^N \log
(|z_n|^2+\epsilon)\quad \nonumber\\
\text{s.t.} \quad
&\|\boldsymbol{y}-\boldsymbol{A}(\boldsymbol{\theta})\boldsymbol{z}\|_2\leq\varepsilon
\label{opt-2}
\end{align}
where $z_n$ denotes the $n$th component of the vector
$\boldsymbol{z}$, and $\epsilon>0$ is a positive parameter to
ensure that the function is well-defined. The optimization
(\ref{opt-2}) can be formulated as an unconstrained optimization
problem by removing the constraint and adding a Tikhonov
regularization term,
$\lambda\|\boldsymbol{y}-\boldsymbol{A}(\boldsymbol{\theta})\boldsymbol{z}\|_2^2$,
to the objective functional, which yields the following
optimization
\begin{align}
\min_{\boldsymbol{z},\boldsymbol{\theta}}\quad
G(\boldsymbol{z},\boldsymbol{\theta})\triangleq &\sum_{n=1}^N \log
(|z_n|^2+\epsilon) +
\lambda\|\boldsymbol{y}-\boldsymbol{A}(\boldsymbol{\theta})\boldsymbol{z}\|_2^2
\nonumber\\
=&
L(\boldsymbol{z})+\lambda\|\boldsymbol{y}-\boldsymbol{A}(\boldsymbol{\theta})\boldsymbol{z}\|_2^2
\label{opt-R4}
\end{align}
where $\lambda>0$ is a regularization parameter controlling the
tradeoff between data fitting and the sparsity of the solution,
and its choice will be more thoroughly discussed later in this
paper.

\section{Proposed Iterative Reweighted Algorithm} \label{sec:algorithm}
We now develop a generalized iterative reweighted $\ell_2$
algorithm for joint dictionary parameter learning and sparse
signal recovery. We resort to a bounded optimization approach,
also known as the majorization-minimization (MM) approach
\cite{LangeHunter00,CandesWakin08}, to solve the optimization
(\ref{opt-R4}). The idea of the MM approach is to iteratively
minimize a simple surrogate function majorizing the given
objective function. Nevertheless, in this paper we will show that
through iteratively decreasing (not necessarily minimizing) the
surrogate function, the iterative process also yields a
non-increasing objective function value and eventually converges
to a stationary point of $G(\boldsymbol{z},\boldsymbol{\theta})$.
To obtain an appropriate surrogate function for (\ref{opt-R4}), we
first find a suitable surrogate function for the log-sum
functional $L(\boldsymbol{z})$. It has been shown in
\cite{FangLi14} that a differentiable and convex surrogate
function majorizing $L(\boldsymbol{z})$ is given by
\begin{align}
Q(\boldsymbol{z}|\boldsymbol{\hat{z}}^{(t)})\triangleq\sum_{n=1}^N
\bigg(\frac{|z_n|^2+\epsilon}{|\hat{z}_n^{(t)}|^2+\epsilon}+\log(|\hat{z}_n^{(t)}|^2+\epsilon)-1
\bigg) \label{surrogate-function}
\end{align}
where $\boldsymbol{\hat{z}}^{(t)}\triangleq
[\hat{z}_1^{(t)}\phantom{0}\ldots\phantom{0}\hat{z}_N^{(t)}]^T$
denotes an estimate of $\boldsymbol{z}$ at iteration $t$. We can
easily verify that
$Q(\boldsymbol{z}|\boldsymbol{\hat{z}}^{(t)})-L(\boldsymbol{z})\geq
0$, with the equality attained when
$\boldsymbol{z}=\boldsymbol{\hat{z}}^{(t)}$. Consequently the
surrogate function for the objective function
$G(\boldsymbol{z},\boldsymbol{\theta})$ is
\begin{align}
S(\boldsymbol{z},\boldsymbol{\theta}|\boldsymbol{\hat{z}}^{(t)})\triangleq
Q(\boldsymbol{z}|\boldsymbol{\hat{z}}^{(t)})+\lambda
\|\boldsymbol{y}-\boldsymbol{A}(\boldsymbol{\theta})\boldsymbol{z}\|_2^2
\label{surrogate-function-R1}
\end{align}



Solving (\ref{opt-R4}) now reduces to minimizing the surrogate
function iteratively. Ignoring terms independent of
$\{\boldsymbol{z},\boldsymbol{\theta}\}$, optimizing the surrogate
function (\ref{surrogate-function-R1}) is simplified as
\begin{align}
\min_{\boldsymbol{z},\boldsymbol{\theta}}\quad &
\boldsymbol{z}^H\boldsymbol{D}^{(t)}\boldsymbol{z} + \lambda
\|\boldsymbol{y}-\boldsymbol{A}(\boldsymbol{\theta})\boldsymbol{z}\|_2^2
\label{opt-R5}
\end{align}
where $[\cdot]^{H}$ denotes the conjugate transpose, and
$\boldsymbol{D}^{(t)}$ is a diagonal matrix given as
\begin{align}
\boldsymbol{D}^{(t)}\triangleq\text{diag}
\bigg\{\frac{1}{|\hat{z}_1^{(t)}|^2+\epsilon},\ldots,\frac{1}{|\hat{z}_N^{(t)}|^2+\epsilon}\bigg\}
\nonumber
\end{align}
Conditioned on $\boldsymbol{\theta}$, the optimal $\boldsymbol{z}$
of (\ref{opt-R5}) can be readily obtained as
\begin{align}
\boldsymbol{z}^{\ast}(\boldsymbol{\theta})=\left(\boldsymbol{A}^H(\boldsymbol{\theta})\boldsymbol{A}(\boldsymbol{\theta})
+\lambda^{-1}\boldsymbol{D}^{(t)}\right)^{-1}\boldsymbol{A}^H(\boldsymbol{\theta})\boldsymbol{y}
\label{eqn-R2}
\end{align}
Substituting (\ref{eqn-R2}) back into (\ref{opt-R5}), the
optimization simply becomes searching for the unknown parameter
$\boldsymbol{\theta}$:
\begin{align}
\min_{\boldsymbol{\theta}}\phantom{0}
f(\boldsymbol{\theta})\triangleq
-\boldsymbol{y}^H\boldsymbol{A}(\boldsymbol{\theta})\left(\boldsymbol{A}^H(\boldsymbol{\theta})\boldsymbol{A}(\boldsymbol{\theta})
+\lambda^{-1}\boldsymbol{D}^{(t)}\right)^{-1}\boldsymbol{A}^H(\boldsymbol{\theta})\boldsymbol{y}
\label{opt-R6}
\end{align}
An analytical solution of the above optimization (\ref{opt-R6}) is
difficult to obtain. Nevertheless, in our algorithm, we only need
to search for a new estimate $\boldsymbol{\hat{\theta}}^{(t+1)}$
such that the following inequality holds
\begin{align}
f(\boldsymbol{\hat{\theta}}^{(t+1)})\leq
f(\boldsymbol{\hat{\theta}}^{(t)}) \label{eqn-R3}
\end{align}
Such an estimate can be easily obtained by using a gradient
descent method. Given $\boldsymbol{\hat{\theta}}^{(t+1)}$,
$\boldsymbol{\hat{z}}^{(t+1)}$ can be obtained via (\ref{eqn-R2}),
with $\boldsymbol{\theta}$ replaced by
$\boldsymbol{\hat{\theta}}^{(t+1)}$, i.e.
\begin{align}
\boldsymbol{\hat{z}}^{(t+1)}=\boldsymbol{z}^{\ast}(\boldsymbol{\hat{\theta}}^{(t+1)})
\label{eqn-R4}
\end{align}

In the following, we show that the new estimate
$\{\boldsymbol{\hat{z}}^{(t+1)},
\boldsymbol{\hat{\theta}}^{(t+1)}\}$ results in a non-increasing
objective function value, that is,
\begin{align}
G(\boldsymbol{\hat{z}}^{(t+1)},\boldsymbol{\hat{\theta}}^{(t+1)})\leq
G(\boldsymbol{\hat{z}}^{(t)},\boldsymbol{\hat{\theta}}^{(t)})
\end{align}
To this goal, we first show the following inequality
\begin{align}
S(\boldsymbol{\hat{z}}^{(t)},\boldsymbol{\hat{\theta}}^{(t)}|\boldsymbol{\hat{z}}^{(t)})
\stackrel{(a)}{\geq}&
S(\boldsymbol{z}^{\ast}(\boldsymbol{\hat{\theta}}^{(t)}),\boldsymbol{\hat{\theta}}^{(t)}
|\boldsymbol{\hat{z}}^{(t)}) \nonumber\\
=& f(\boldsymbol{\hat{\theta}}^{(t)}) +\text{constant} \nonumber\\
\stackrel{(b)}{\geq} & f(\boldsymbol{\hat{\theta}}^{(t+1)}) +\text{constant} \nonumber\\
=&
S(\boldsymbol{z}^{\ast}(\boldsymbol{\hat{\theta}}^{(t+1)}),\boldsymbol{\hat{\theta}}^{(t+1)}|\boldsymbol{\hat{z}}^{(t)})
\nonumber\\
\stackrel{(c)}{=}&
S(\boldsymbol{\hat{z}}^{(t+1)},\boldsymbol{\hat{\theta}}^{(t+1)}|\boldsymbol{\hat{z}}^{(t)})
\label{inequality-R1}
\end{align}
where $(a)$ comes from the fact that
$\boldsymbol{z}^{\ast}(\boldsymbol{\theta})$ is the optimal
solution to the optimization (\ref{opt-R5}); $(b)$ and $(c)$
follow from (\ref{eqn-R3}) and (\ref{eqn-R4}), respectively.
Moreover, we have
\begin{align}
&G(\boldsymbol{\hat{z}}^{(t+1)},\boldsymbol{\hat{\theta}}^{(t+1)})-
S(\boldsymbol{\hat{z}}^{(t+1)},\boldsymbol{\hat{\theta}}^{(t+1)}|\boldsymbol{\hat{z}}^{(t)})
\nonumber\\
=&
L(\boldsymbol{\hat{z}}^{(t+1)})-Q(\boldsymbol{\hat{z}}^{(t+1)}|\boldsymbol{\hat{z}}^{(t)})
\nonumber\\
\stackrel{(a)}{\leq}&
L(\boldsymbol{\hat{z}}^{(t)})-Q(\boldsymbol{\hat{z}}^{(t)}|\boldsymbol{\hat{z}}^{(t)})
\nonumber\\
=& G(\boldsymbol{\hat{z}}^{(t)},\boldsymbol{\hat{\theta}}^{(t)})-
S(\boldsymbol{\hat{z}}^{(t)},\boldsymbol{\hat{\theta}}^{(t)}|\boldsymbol{\hat{z}}^{(t)})
\label{inequality-R2}
\end{align}
where $(a)$ follows from the fact that
$Q(\boldsymbol{z}|\boldsymbol{\hat{z}}^{(t)})-L(\boldsymbol{z})$
attains its minimum when
$\boldsymbol{z}=\boldsymbol{\hat{z}}^{(t)}$. Combining
(\ref{inequality-R1})--(\ref{inequality-R2}), we eventually arrive
at
\begin{align}
G(\boldsymbol{\hat{z}}^{(t+1)},\boldsymbol{\hat{\theta}}^{(t+1)})=&
G(\boldsymbol{\hat{z}}^{(t+1)},\boldsymbol{\hat{\theta}}^{(t+1)})-
S(\boldsymbol{\hat{z}}^{(t+1)},\boldsymbol{\hat{\theta}}^{(t+1)}|\boldsymbol{\hat{z}}^{(t)})
\nonumber\\
&+
S(\boldsymbol{\hat{z}}^{(t+1)},\boldsymbol{\hat{\theta}}^{(t+1)}|\boldsymbol{\hat{z}}^{(t)})
\nonumber\\
\leq&
G(\boldsymbol{\hat{z}}^{(t)},\boldsymbol{\hat{\theta}}^{(t)})-
S(\boldsymbol{\hat{z}}^{(t)},\boldsymbol{\hat{\theta}}^{(t)}|\boldsymbol{\hat{z}}^{(t)})
\nonumber\\
&+S(\boldsymbol{\hat{z}}^{(t+1)},\boldsymbol{\hat{\theta}}^{(t+1)}|\boldsymbol{\hat{z}}^{(t)})
\nonumber\\
\leq&
G(\boldsymbol{\hat{z}}^{(t)},\boldsymbol{\hat{\theta}}^{(t)})
\label{inequality-R3}
\end{align}
We see that through iteratively decreasing (not necessarily
minimizing) the surrogate function, the objective function
$G(\boldsymbol{z},\boldsymbol{\theta})$ is guaranteed to be
non-increasing at each iteration.

For clarification, we summarize our algorithm as follows.

\begin{center}
\textbf{Iterative Reweighted Algorithm I}
\end{center}

\vspace{0cm} \noindent
\begin{tabular}{lp{7.7cm}}
\hline 1.& Given an initialization $\boldsymbol{\hat{z}}^{(0)},
\boldsymbol{\hat{\theta}}^{(0)}$, and
a pre-selected regularization parameter $\lambda$.\\
2.& At iteration $t=0,1,\ldots$: Based on the estimate
$\boldsymbol{\hat{z}}^{(t)}$, construct the surrogate function as
depicted in (\ref{surrogate-function-R1}). Search for a new
estimate of the unknown parameter vector, denoted as
$\boldsymbol{\hat{\theta}}^{(t+1)}$, by using the gradient descent
method such that the inequality (\ref{eqn-R3}) is satisfied.
Compute a new estimate of the sparse signal, denoted as
$\boldsymbol{\hat{z}}^{(t+1)}$, via (\ref{eqn-R4}).\\
3.& Go to Step 2 if
$\|\boldsymbol{\hat{z}}^{(t+1)}-\boldsymbol{\hat{z}}^{(t)}\|_2>\varepsilon$,
where $\varepsilon$ is a prescribed tolerance value; otherwise
stop.\\
\hline
\end{tabular}

\vspace{0.3cm}



We see that in our algorithm, the unknown parameters and the
signal are refined in a gradual and interweaved manner. This
enables the algorithm, with a high probability, comes to a
reasonably nearby point during the first few iterations, and
eventually converges to the correct basin of attraction. In
addition, similar to \cite{ChartrandYin08}, the parameter
$\epsilon$ used throughout our optimization can be gradually
decreased instead of remaining fixed. For example, at the
beginning, $\epsilon$ can be set to a relatively large value, say
1. We then gradually reduce the value of $\epsilon$ in the
subsequent iterations until $\epsilon$ attains a sufficiently
small value, e.g. $10^{-8}$. Numerical results demonstrate that
this gradual refinement of the parameter $\epsilon$ can further
improve the probability of finding the correct solution.

The second step of the proposed algorithm involves searching for a
new estimate of the unknown parameter vector to meet the condition
(\ref{eqn-R3}). As mentioned earlier, this can be accomplished via
a gradient-based search algorithm. Details of computing the
gradient of $f(\boldsymbol{\theta})$ with respect to
$\boldsymbol{\theta}$ are provided in Appendix \ref{appA}. Also,
to achieve a better reconstruction accuracy, the estimates of
$\{\theta_i\}$ can be refined in a sequential manner. Our
experiments suggest that a new estimate which satisfies
(\ref{eqn-R3}) can be easily obtained within only a few
iterations.




The main computational task of our proposed algorithm at each
iteration is to calculate
$\boldsymbol{z}^{\ast}(\boldsymbol{\theta})$ (as per
(\ref{eqn-R2})) and the first derivative of
$f(\boldsymbol{\theta})$ with respect to $\boldsymbol{\theta}$,
both of which involve computing the inverse of the following
$N\times N$ matrix:
$\boldsymbol{A}^H(\boldsymbol{\theta})\boldsymbol{A}(\boldsymbol{\theta})
+\lambda^{-1}\boldsymbol{D}^{(t)}$. By using the Woodbury
identity, this $N\times N$ matrix inversion can be converted to an
$M\times M$ matrix inversion (this conversion is meaningful when
$M\ll N$). The computational complexity of the proposed method can
be further reduced by introducing a pruning operation, that is, at
each iteration, we prune those small coefficients along with their
associated frequency components such that the dimensions of the
signal $\boldsymbol{z}$ and the parameter $\boldsymbol{\theta}$
keep shrinking as the iterative process evolves, eventually
retaining only a few prominent nonzero coefficients. A hard
thresholding rule can be used to prune those irrelevant frequency
components. Specifically, if the coefficient $\hat{z}_n^{(t)}$ is
less than a pre-specified small value $\tau$, i.e.
$\hat{z}_n^{(t)}\leq\tau$, then the associated frequency component
$\hat{\theta}_n^{(t)}$ can be removed from further consideration
since its contribution to the signal synthesis is negligible.

Note that the above pruning procedure cannot be applied to our
previous algorithm \cite{FangLi14} developed for the scenario of
noiseless measurements. To see this, the previous algorithm
requires the computation of the inverse of the following $M\times
M$ matrix
$\boldsymbol{A}(\boldsymbol{\theta})(\boldsymbol{D}^{(t)})^{-1}\boldsymbol{A}^H(\boldsymbol{\theta})$
at each iteration. Performing pruning operations will result in an
ill-posed inverse problem since the above matrix will eventually
become singular as the dimension of
$\boldsymbol{A}(\boldsymbol{\theta})$ shrinks. The proposed method
in the current work is therefore computationally more attractive
than our previous algorithm, particularly when the number of
observed data samples, $M$, is large. Note that the proposed
method can also be used to solve the noiseless problem by
adaptively updating the regularization parameter $\lambda$.
Details of how to adaptively update $\lambda$ is discussed next.










\section{Adaptive Update of $\lambda$}
\label{sec:lambda-choice} As mentioned earlier, $\lambda$ is a
regularization parameter controlling the tradeoff between the
sparsity of the solution and the data fitting error. Clearly, a
small $\lambda$ leads to a sparse solution, whereas a larger
$\lambda$ renders a less sparse but better-fitting solution. As a
consequence, in scenarios where frequency components are
closely-spaced, choosing a small $\lambda$ may result in an
underestimation of the frequency components while an excessively
large $\lambda$ may lead to an overestimation. Thus the choice of
$\lambda$ is critical to the recovery performance.

When the knowledge of the noise level is known \emph{a priori},
the regularization parameter $\lambda$ can be chosen such that the
norm of the residual matches the noise level of the data. This
selection rule is also known as the discrepancy principle
\cite{BauerLukas11}. For the case of unknown noise variance, the
L-curve method has been shown to provide a reasonably good and
robust parameter choice \cite{BauerLukas11} in some experiments.
Nevertheless, the L-curve method is computationally expensive for
our case since, in order to plot the L-curve, it requires us to
solve the optimization problem (\ref{opt-R4}) for a number of
different values of $\lambda$. To the best of our knowledge, a
general rule for regularization parameter selection remains an
open issue. In this section, we propose a simple yet effective
scheme for adaptively updating the parameter $\lambda$ during the
iterative process. The developed scheme does not require the
knowledge of the noise variance.







Note that iterative reweighted methods have a close connection
with sparse Bayesian learning algorithms
\cite{Tipping01,JiXue08,FangShen15}. In fact, a dual-form analysis
\cite{WipfNagaranjan10} reveals that sparse Bayesian learning can
be considered as a non-separable iterative reweighted strategy
solving a non-separable penalty function. Inspired by this
insight, it is expected the mechanism inherent in the sparse
Bayesian learning method to achieve automatic balance between the
sparsity and the fitting error should also work for the iterative
reweighted methods. Let us first briefly examine how the sparse
Bayesian learning algorithm works. In the sparse Bayesian learning
framework, the observation noise is assumed to be white Gaussian
noise with zero mean and variance $\delta\triangleq\sigma^2$, and
the sparse signal $\boldsymbol{z}$ is assigned a Gaussian prior
distribution \cite{Tipping01}
\begin{align}
p(\boldsymbol{z}|\boldsymbol{\alpha})=\prod_{n=1}^N
p(z_n|\alpha_n) \nonumber
\end{align}
where $p(z_n|\alpha_n)=\mathcal{N}(z_n|0,\alpha_n^{-1})$ and
$\boldsymbol{\alpha}\triangleq\{\alpha_n\}$. Here each $\alpha_n$
is the inverse variance (precision) of the Gaussian distribution,
and a non-negative sparsity-controlling hyperparameter. For each
iteration, given a set of estimated hyperparameters
$\{\alpha_n^{(t)}\}$, the maximum a posterior (MAP) estimator of
$\boldsymbol{z}$ can be obtained via
\begin{align}
\boldsymbol{\hat{z}}^{(t)}=\arg\min_{\boldsymbol{z}} \quad
\boldsymbol{z}^H\boldsymbol{D}^{(t)}\boldsymbol{z} + \delta^{-1}
\|\boldsymbol{y}-\boldsymbol{A}(\boldsymbol{\theta})\boldsymbol{z}\|_2^2
\label{opt-R7}
\end{align}
where
$\boldsymbol{D}^{(t)}\triangleq\text{diag}(\alpha_1^{(t)},\ldots,\alpha_N^{(t)})$.
Meanwhile, given the estimated sparse signal
$\boldsymbol{\hat{z}}^{(t)}$ and its posterior covariance matrix,
the hyperparameters $\{\alpha_i\}$ are re-estimated. In this
Bayesian framework, the tradeoff between the sparsity and the data
fitting is automatically achieved by employing a probabilistic
model for the sparse signal $\boldsymbol{z}$, and the tradeoff
tuning parameter is equal to the inverse of the noise variance
$\delta$ (cf. (\ref{opt-R7})).

Comparing (\ref{opt-R5}) and (\ref{opt-R7}), we see that the
sparse Bayesian learning method is similar to our proposed
iterative reweighted algorithm, except that the reweighted
diagonal matrix $\boldsymbol{D}^{(t)}$ is updated in different
ways, and the sparse Bayesian learning method assumes the
dictionary $\boldsymbol{A}(\boldsymbol{\theta})$ is fully known
and, therefore, does not involve the optimization of the
dictionary parameter $\boldsymbol{\theta}$. Following
(\ref{opt-R7}), an appropriate choice of $\lambda$ in
(\ref{opt-R5}) is to make it inversely proportional to the noise
variance, i.e. $\lambda=d\delta^{-1}$, where $d$ is a constant
scaling factor. Note that when the noise variance is unknown
\emph{a priori}, the noise variance $\delta$ can be iteratively
estimated, based on which the tuning parameter $\lambda$ can be
iteratively updated. A reasonable estimate of the noise variance
is given by
\begin{align}
\hat{\delta}^{(t)}=\frac{\|\boldsymbol{y}-
\boldsymbol{A}(\boldsymbol{\hat{\theta}}^{(t)})\boldsymbol{\hat{z}}^{(t)}\|_2^2}{M}
\end{align}
and accordingly $\lambda^{(t)}$ can be updated as
\begin{align}
\lambda^{(t)}=\frac{d}{\hat{\delta}^{(t)}}=\frac{dM}{\|\boldsymbol{y}-
\boldsymbol{A}(\boldsymbol{\hat{\theta}}^{(t)})\boldsymbol{\hat{z}}^{(t)}\|_2^2}
\label{eqn-R5}
\end{align}


The iterative update of $\lambda$ can be seamlessly integrated
into our algorithm, which is summarized as follows.

\begin{center}
\textbf{Iterative Reweighted Algorithm II}
\end{center}

\vspace{0cm} \noindent
\begin{tabular}{lp{7.7cm}}
\hline 1.& Given an initialization $\boldsymbol{\hat{z}}^{(0)},
\boldsymbol{\hat{\theta}}^{(0)}$, and
$\lambda^{(0)}$.\\
2.& At iteration $t=0,1,\ldots$: Based on
$\boldsymbol{\hat{z}}^{(t)}$ and $\lambda^{(t)}$, construct the
surrogate function as depicted in (\ref{surrogate-function-R1}).
Search for a new estimate of the unknown parameter vector, denoted
as $\boldsymbol{\hat{\theta}}^{(t+1)}$, by using the gradient
descent method such that the inequality (\ref{eqn-R3}) is
satisfied. Compute a new estimate of the sparse signal, denoted as
$\boldsymbol{\hat{z}}^{(t+1)}$, via (\ref{eqn-R4}). Compute a new
regularization
parameter $\lambda^{(t+1)}$ according to (\ref{eqn-R5}).\\
3.& Go to Step 2 if
$\|\boldsymbol{\hat{z}}^{(t+1)}-\boldsymbol{\hat{z}}^{(t)}\|_2>\varepsilon$,
where $\varepsilon$ is a prescribed tolerance value; otherwise
stop.\\
\hline
\end{tabular}

\vspace{0.3cm}


The above algorithm, in fact,  can be interpreted as solving the
following optimization problem
\begin{align}
\min_{\boldsymbol{z},\boldsymbol{\theta},\lambda}
\tilde{G}(\boldsymbol{z},\boldsymbol{\theta},\lambda)\triangleq
&\sum_{n=1}^N \log (|z_n|^2+\epsilon) +
\lambda\|\boldsymbol{y}-\boldsymbol{A}(\boldsymbol{\theta})\boldsymbol{z}\|_2^2-dM\log\lambda
\nonumber\\
=&
L(\boldsymbol{z})+\lambda\|\boldsymbol{y}-\boldsymbol{A}(\boldsymbol{\theta})\boldsymbol{z}\|_2^2-dM\log\lambda
\label{opt-R8}
\end{align}
To see this, note that given an estimate of
$\{\boldsymbol{\hat{z}}^{(t)}, \boldsymbol{\hat{\theta}}^{(t)}\}$,
the optimal $\lambda$ of (\ref{opt-R8}) is given by
(\ref{eqn-R5}). On the other hand, for a fixed $\lambda^{(t)}$,
the above optimization reduces to (\ref{opt-R4}), in which case a
new estimate
$\{\boldsymbol{\hat{z}}^{(t+1)},\boldsymbol{\hat{\theta}}^{(t+1)}\}$
can be obtained according to (\ref{eqn-R3}) and (\ref{eqn-R4}).
Therefore the proposed algorithm ensures that the objective
function value of (\ref{opt-R8}) is non-increasing at each
iteration, and eventually converges to a stationary point of
(\ref{opt-R8}). The last term, $dm\log\lambda$, in (\ref{opt-R8})
is a regularization term included to pull $\lambda$ away from
zero. Without this term, the optimization (\ref{opt-R8}) becomes
meaningless since the optimal $\lambda$ in this case is equal to
zero.








\section{Extension To The MMV Model} \label{sec:mmv}
In some practical applications such as EEG/MEG source localization
and DOA estimation, multiple measurements
$\{\boldsymbol{y}_1,\ldots,\boldsymbol{y}_L\}$ of a time series
process may be available. This motivates us to consider the
super-resolution compressed sensing problem in a multiple
measurement vector (MMV) framework \cite{CotterRao05}
\begin{align}
\boldsymbol{Y}=\boldsymbol{A}(\boldsymbol{\theta})\boldsymbol{Z}+\boldsymbol{W}
\end{align}
where
$\boldsymbol{Y}\triangleq[\boldsymbol{y}_1\phantom{0}\boldsymbol{y}_2\cdots\boldsymbol{y}_L]$
is an observation matrix consisting of $L$ observed vectors,
$\boldsymbol{Z}\triangleq[\boldsymbol{z}_1\phantom{0}\boldsymbol{z}_2\cdots\boldsymbol{z}_L]$
is a sparse matrix with each row representing a possible source or
frequency component, and $\boldsymbol{W}$ denotes the noise
matrix. Note that in the MMV model, we assume that the support of
the sparse signal remains unchanged over time, that is, the matrix
$\boldsymbol{Z}$ has a common row sparsity pattern. This is a
reasonable assumption in many applications where the variations of
locations or frequencies are slow compared to the sampling rate.
The problem of joint dictionary parameter learning and sparse
signal recovery can be formulated as follows
\begin{align}
\min_{\boldsymbol{Z},\boldsymbol{\theta}}\quad &
\|\boldsymbol{u}\|_{0} \nonumber\\
\text{s.t.} \quad
&\|\boldsymbol{Y}-\boldsymbol{A}(\boldsymbol{\theta})\boldsymbol{Z}\|_F\leq\varepsilon
\label{opt-R9}
\end{align}
where $\|\boldsymbol{X}\|_F$ denotes the Frobenius norm of the
matrix $\boldsymbol{X}$, and $\boldsymbol{u}$ is a column vector
with its entry $u_n$ defined as
\begin{align}
u_n\triangleq\|\boldsymbol{z}_{n\cdot}\|_2 \quad \forall
n=1,\ldots,N \nonumber
\end{align}
in which $\boldsymbol{z}_{n\cdot}$ represents the $n$th row of
$\boldsymbol{Z}$. Thus $\|\boldsymbol{u}\|_{0}$ equals to the
number of nonzero rows in $\boldsymbol{Z}$. Clearly, the
optimization (\ref{opt-R9}) aims to search for a set of unknown
parameters $\{\theta_n\}$ with which the observed matrix
$\boldsymbol{Y}$ can be represented by as few atoms as possible
with a specified error tolerance. Again, to make the problem
(\ref{opt-R9}) tractable, the $\ell_0$-norm can be replaced with
the log-sum functional, which leads to the following optimization
\begin{align}
\min_{\boldsymbol{Z},\boldsymbol{\theta}}\quad &
L(\boldsymbol{Z})=\sum_{n=1}^N \log
(\|\boldsymbol{z}_{n\cdot}\|_2^2+\epsilon)\quad \nonumber\\
\text{s.t.} \quad
&\|\boldsymbol{Y}-\boldsymbol{A}(\boldsymbol{\theta})\boldsymbol{Z}\|_F\leq\varepsilon
\label{opt-R10}
\end{align}
The constraint in the above optimization can be absorbed into the
objective function as a Tikhonov regularization term and
(\ref{opt-R10}) can be rewritten as
\begin{align}
\min_{\boldsymbol{Z},\boldsymbol{\theta}}\quad
G(\boldsymbol{Z},\boldsymbol{\theta})\triangleq L(\boldsymbol{Z})
+ \lambda
\|\boldsymbol{Y}-\boldsymbol{A}(\boldsymbol{\theta})\boldsymbol{Z}\|_F^2
\label{opt-R11}
\end{align}
Again, we resort to the majorization-minimization (MM) approach to
solve (\ref{opt-R11}). It can be readily verified that a suitable
surrogate function majorizing the log-sum functional
$L(\boldsymbol{Z})$ is given by
\begin{align}
Q(\boldsymbol{Z}|\boldsymbol{\hat{Z}}^{(t)})\triangleq
\sum_{n=1}^N
\left(\frac{\|\boldsymbol{z}_{n\cdot}\|_2^2+\epsilon}{\|\boldsymbol{\hat{z}}_{n\cdot}^{(t)}\|_2^2+\epsilon}
+\log
(\|\boldsymbol{\hat{z}}_{n\cdot}^{(t)}\|_2^2+\epsilon)-1\right)
\end{align}
Defining
\begin{align}
\boldsymbol{D}^{(t)}\triangleq\text{diag}\left(\frac{1}{\|\boldsymbol{\hat{z}}_{1\cdot}^{(t)}\|_2^2+\epsilon}
,\ldots,\frac{1}{\|\boldsymbol{\hat{z}}_{N\cdot}^{(t)}\|_2^2+\epsilon}\right)
\nonumber
\end{align}
and ignoring terms independent of
$\{\boldsymbol{Z},\boldsymbol{\theta}\}$, optimizing
(\ref{opt-R11}) becomes iteratively minimizing the following
surrogate function
\begin{align}
\min_{\boldsymbol{Z},\boldsymbol{\theta}} \quad
\text{tr}\left(\boldsymbol{Z}^H\boldsymbol{D}^{(t)}\boldsymbol{Z}\right)+\lambda
\|\boldsymbol{Y}-\boldsymbol{A}(\boldsymbol{\theta})\boldsymbol{Z}\|_F^2
\label{opt-R12}
\end{align}
Given $\boldsymbol{\theta}$ fixed, the optimal $\boldsymbol{Z}$ of
(\ref{opt-R12}) can be readily obtained as
\begin{align}
\boldsymbol{Z}^{\ast}(\boldsymbol{\theta})=\left(\boldsymbol{A}^H(\boldsymbol{\theta})\boldsymbol{A}(\boldsymbol{\theta})
+\lambda^{-1}\boldsymbol{D}^{(t)}\right)^{-1}\boldsymbol{A}^H(\boldsymbol{\theta})\boldsymbol{Y}
\label{eqn-R6}
\end{align}
Substituting (\ref{eqn-R6}) back into (\ref{opt-R12}), the
optimization simply becomes searching for the unknown parameter
$\boldsymbol{\theta}$:
\begin{align}
&\min_{\boldsymbol{\theta}}\phantom{0}  f(\boldsymbol{\theta})
\nonumber\\
&\triangleq -\text{tr}\left\{
\boldsymbol{Y}^H\boldsymbol{A}(\boldsymbol{\theta})\left(\boldsymbol{A}^H(\boldsymbol{\theta})\boldsymbol{A}(\boldsymbol{\theta})
+\lambda^{-1}\boldsymbol{D}^{(t)}\right)^{-1}\boldsymbol{A}^H(\boldsymbol{\theta})\boldsymbol{Y}\right\}
\end{align}
Again, in our algorithm, we only need to search for a new estimate
$\boldsymbol{\hat{\theta}}^{(t+1)}$ such that the following
inequality holds valid
\begin{align}
f(\boldsymbol{\hat{\theta}}^{(t+1)})\leq
f(\boldsymbol{\hat{\theta}}^{(t)})
\end{align}
Such an estimate can be found by using a gradient-based search
algorithm. The derivative of $f(\boldsymbol{\theta})$ with respect
to $\boldsymbol{\theta}$ is similar to that in Appendix
\ref{appA}, except with $\boldsymbol{y}$ replaced by
$\boldsymbol{Y}$. Given $\boldsymbol{\hat{\theta}}^{(t+1)}$,
$\boldsymbol{\hat{z}}^{(t+1)}$ can be obtained via (\ref{eqn-R6}),
with $\boldsymbol{\theta}$ replaced by
$\boldsymbol{\hat{\theta}}^{(t+1)}$. Following an analysis similar
to (\ref{inequality-R1})--(\ref{inequality-R3}), we can show that
the estimate
$\{\boldsymbol{\hat{Z}}^{(t+1)},\boldsymbol{\hat{\theta}}^{(t+1)}\}$
results in a non-increasing objective function value, that is,
$G(\boldsymbol{\hat{Z}}^{(t+1)},\boldsymbol{\hat{\theta}}^{(t+1)})\leq
G(\boldsymbol{\hat{Z}}^{(t)},\boldsymbol{\hat{\theta}}^{(t)})$.
Therefore the proposed algorithm is guaranteed to converge to a
stationary point of $G(\boldsymbol{Z},\boldsymbol{\theta})$.








\section{Exact Reconstruction Analysis} \label{sec:analysis}
In this section, we provide an insightful analysis of
(\ref{opt-2}) to shed light on conditions under which exact
reconstruction is possible. We assume the noiseless case since
exact recovery is impossible when noise is present. For the
noiseless case, the optimization (\ref{opt-2}) simply becomes
\begin{align}
\min_{\boldsymbol{z},\boldsymbol{\theta}}\quad &
L(\boldsymbol{z})=\sum_{i=1}^N \log
(|z_i|^2+\epsilon)\quad \nonumber\\
\text{s.t.} \quad
&\boldsymbol{y}=\boldsymbol{A}(\boldsymbol{\theta})\boldsymbol{z}
\label{opt-R1}
\end{align}
Note that an iterative reweighted algorithm was developed in our
earlier work \cite{FangLi14} to solve (\ref{opt-R1}), and achieves
superior exact recovery performance. Conducting a rigorous
theoretical analysis of (\ref{opt-R1}), however, is difficult. We,
instead, consider an alternative optimization that is more amiable
for our analysis. It was shown in \cite{RaoDelgado99} that the
log-sum function defined in (\ref{opt-R1}) behaves like the
$\ell_0$-norm when $\epsilon$ is sufficiently small. Particularly,
when $\epsilon=0$, the log-sum function is essentially the same as
the $\ell_0$-norm. To gain insight into (\ref{opt-R1}), we examine
the exact reconstruction condition of the following optimization
\begin{align}
\min_{\boldsymbol{z},\boldsymbol{\theta}}\quad &
\|\boldsymbol{z}\|_0 \nonumber\\
\text{s.t.} \quad
&\boldsymbol{y}=\boldsymbol{A}(\boldsymbol{\theta})\boldsymbol{z}
\label{opt-R2}
\end{align}
Let $\{\boldsymbol{\theta}_0, \boldsymbol{z}_0\}$ and
$\{\boldsymbol{\theta}^{\ast}, \boldsymbol{z}^{\ast}\}$ denote the
groundtruth and the globally optimal solution to (\ref{opt-R2}),
respectively. In addition, define $\boldsymbol{\alpha}_0$ as a
$K$-dimensional vector obtained by retaining only nonzero
coefficients of $\boldsymbol{z}_0$, and $\boldsymbol{\omega}_0$ is
a $K$-dimensional parameter vector obtained by keeping those
corresponding entries in $\boldsymbol{\theta}_0$. Similarly, we
obtain $\{\boldsymbol{\alpha}^{\ast},\boldsymbol{\omega}^{\ast}\}$
from $\{\boldsymbol{\theta}^{\ast}, \boldsymbol{z}^{\ast}\}$. We
now show under what condition the global solution of
(\ref{opt-R2}) equals to the groundtruth. We proceed by
contradiction. Assume that the globally optimal solution does not
coincide with the groundtruth, i.e.
$\{\boldsymbol{\alpha}_0,\boldsymbol{\omega}_0\}\neq\{\boldsymbol{\alpha}^{\ast},\boldsymbol{\omega}^{\ast}\}$.
Then we have
\begin{align}
&\boldsymbol{A}(\boldsymbol{\omega}_0)\boldsymbol{\alpha}_0=
\boldsymbol{A}(\boldsymbol{\omega}^{\ast})\boldsymbol{\alpha}^{\ast}
\nonumber\\
\Rightarrow &
\left[\boldsymbol{A}(\boldsymbol{\omega}_0)\phantom{0}\boldsymbol{A}(\boldsymbol{\omega}^{\ast})\right]
\left[
\begin{array}{c}\boldsymbol{\alpha}_0 \\ -\boldsymbol{\alpha}^{\ast}\end{array}
\right]=\boldsymbol{0} \label{eqn-R1}
\end{align}
Since $\boldsymbol{z}^{\ast}$ is the solution of (\ref{opt-R2}),
we have $\|\boldsymbol{z}^{\ast}\|_0\leq K$. Thus the matrix
$[\boldsymbol{A}(\boldsymbol{\omega}_0)\phantom{0}\boldsymbol{A}(\boldsymbol{\omega}^{\ast})]$
has at most $2K$ non-identical columns, with each column
characterized by a distinct parameter $\omega$. Note that without
loss of generality, we assume that the two sets
$\boldsymbol{\omega}_0$ and $\boldsymbol{\omega}^{\ast}$ do not
share any identical frequency components. Otherwise, the
repetitive components (columns) can be removed. Define
$\boldsymbol{\bar{\omega}}\triangleq\{\boldsymbol{\omega}_0,
\boldsymbol{\omega}^{\ast}\}$, we can write
\begin{align}
\boldsymbol{A}(\boldsymbol{\bar{\omega}})\triangleq
\left[\boldsymbol{A}(\boldsymbol{\omega}_0)\phantom{0}\boldsymbol{A}(\boldsymbol{\omega}^{\ast})\right]
\nonumber
\end{align}
Clearly, $\boldsymbol{A}(\boldsymbol{\bar{\omega}})$ is a
Vandermonde matrix. Since all frequency components in the set
$\boldsymbol{\bar{\omega}}$ are distinct, the matrix
$\boldsymbol{A}(\boldsymbol{\bar{\omega}})$ is full column rank
when $M\geq 2K$, in which case there does not exist any nonzero
vector to satisfy (\ref{eqn-R1}). Therefore given $M\geq 2K$, we
should reach that
$\{\boldsymbol{\alpha}_0,\boldsymbol{\omega}_0\}=\{\boldsymbol{\alpha}^{\ast},\boldsymbol{\omega}^{\ast}\}$,
i.e. solving (\ref{opt-R2}) yields the exact solution. When
$\epsilon$ in (\ref{opt-R1}) approaches zero, i.e.
$\epsilon\rightarrow 0$, the global solutions of (\ref{opt-R1})
and (\ref{opt-R2}) coincide. Hence the global solution of
(\ref{opt-R1}) also provides an exact recovery when $M\geq 2K$.




\section{Simulation Results} \label{sec:simulation}
We now carry out experiments to illustrate the performance of our
proposed super-resolution iterative reweighted $\ell_2$ algorithm
(referred to as SURE-IR)\footnote{Codes are available at
http://www.junfang-uestc.net/codes/Sure-IR.rar}. In our
simulations, the initial value of $\lambda$ and the pruning
threshold $\tau$ are set equal to $\lambda^{(0)}=0.01$ and
$\tau=0.05$, respectively. Also, to improve the stability of our
proposed algorithm, the initial value of $\lambda$ is kept
unchanged and the frequency components are unpruned during the
first few iterations. The parameter $d$ used in (\ref{eqn-R5}) to
update $\lambda$ is set to $d=5$. We compare our proposed
algorithm with other existing state-of-the-art super-resolution
compressed sensing methods, namely, the Bayesian dictionary
refinement compressed sensing algorithm (denoted as DicRefCS)
\cite{HuShi12}, the root-MUSIC based spectral iterative hard
thresholding (SIHT) \cite{DuarteBaraniuk13}, the atomic norm
minimization via the semi-definite programming (SDP) approach
\cite{TangBhaskar12,TangBhaskar13}, and the off-grid sparse
Bayesien inference (OGSBI) algorithm \cite{YangXie13}. Among these
methods, the SURE-IR, DicRefCS, and the OGSBI methods require to
pre-specify the initial grid points. In our experiments, the
initial grid points are set to be
$\boldsymbol{\theta}^{(0)}=(2\pi/N)[0\phantom{0}\ldots\phantom{0}N-1]^T$,
where we choose $N=64$ for the SURE-IR and the DicRefCS methods.
While for the OGSBI method, a much finer grid ($N=200$) is used to
improve the Taylor approximation accuracy and the recovery
performance.


In our experiments, the signal $\boldsymbol{y}_T\triangleq
[y_1\phantom{0}\ldots\phantom{0}y_T]^T$ is a mixture of $K$
complex sinusoids corrupted by independent and identically
distributed (i.i.d.) zero-mean Gaussian noise, i.e.
\begin{align}
y_l=\sum_{k=1}^K \alpha_k e^{-j\omega_k l} +w_l \qquad l=1,\ldots,
T \nonumber
\end{align}
where the frequencies $\{\omega_k\}$ are uniformly generated over
$[0,2\pi)$ and the amplitudes $\{\alpha_k\}$ are uniformly
distributed on the unit circle. The measurements $\boldsymbol{y}$
are obtained by randomly selecting $M$ entries from $T$ elements
of $\boldsymbol{y}_T$. The observation quality is measured by the
peak-signal-to-noise ratio (PSNR) which is defined as
$\text{PSNR}\triangleq 10\log_{10}(1/\sigma^2)$, where $\sigma^2$
denotes the noise variance.

We introduce two metrics to evaluate the recovery performance of
respective algorithms, namely, the reconstruction signal-to-noise
ratio (RSNR) and the success rate. The RSNR measures the accuracy
of reconstructing the original signal $\boldsymbol{y}_T$ from the
partial observations $\boldsymbol{y}$, and is defined as
\begin{align}
\text{RSNR}=20\log_{10}\left(\frac{\|\boldsymbol{y}_T\|_2}{\|\boldsymbol{y}_T-
\boldsymbol{\hat{y}}_T\|_2}\right) \nonumber
\end{align}
The other metric evaluates the success rate of exactly resolving
the $K$ frequency components $\{\omega_k\}$. The success rate is
computed as the ratio of the number of successful trials to the
total number of independent runs, where $\{\alpha_k\}$,
$\{\omega_k\}$ and the sampling indices (used to obtain
$\boldsymbol{y}$) are randomly generated for each run. A trial is
considered successful if the number of frequency components is
estimated correctly and the estimation error between the estimated
frequencies $\{\hat{\omega}_k\}$ and the true parameters
$\{\omega_k\}$ is smaller than $10^{-3}$, i.e.
$\frac{1}{2\pi}\|\boldsymbol{\omega}-\boldsymbol{\hat{\omega}}\|_2\leq
10^{-3}$. Note that the SIHT and the SDP methods require the
knowledge of the number of complex sinusoids, $K$, which is
assumed perfectly known to them. The OGSBI method usually results
in an overestimated solution which may contain multiple peaks
around each true frequency component. To compute the success rate
for the OGSBI, we only keep those $K$ frequency components
associated with the first $K$ largest coefficients.

\begin{figure*}[!t]
 \centering
\subfigure[RSNRs vs. $M$.]{\includegraphics[width=9cm]{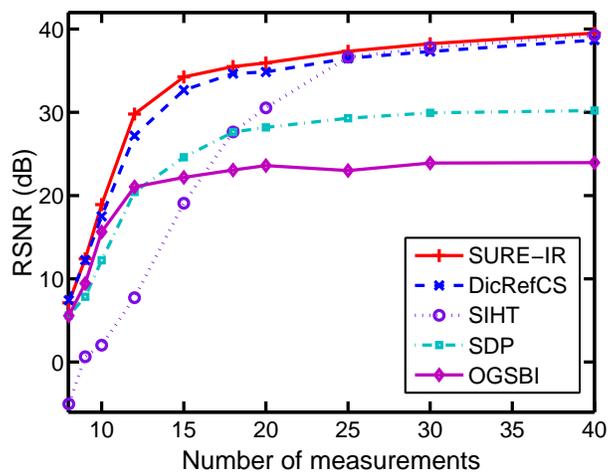}}
 \hfil
\subfigure[Success rates vs.
$M$]{\includegraphics[width=9cm]{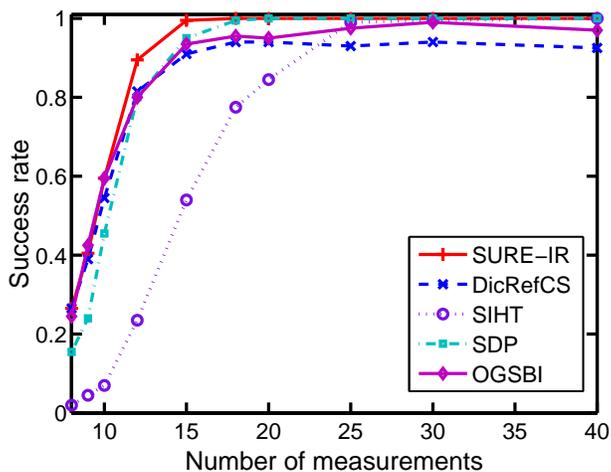}}
  \caption{RSNRs and success rates of respective algorithms vs. $M$, $T=64$, $K=3$, and
$\text{PSNR}=25\text{dB}$.}
   \label{fig1}
\end{figure*}

\begin{figure*}[!t]
 \centering
\subfigure[RSNRs vs. $K$.]{\includegraphics[width=9cm]{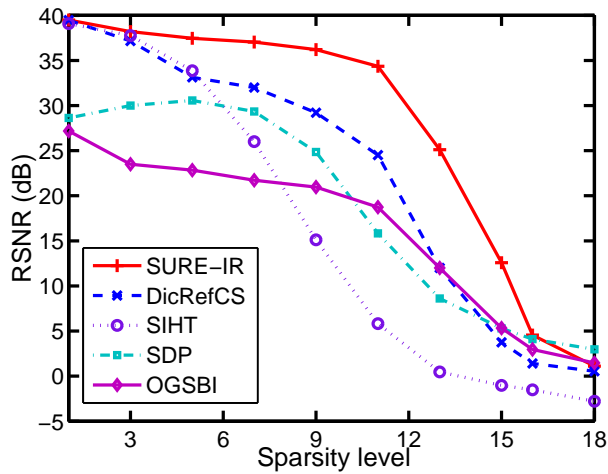}}
 \hfil
\subfigure[Success rates vs.
$K$]{\includegraphics[width=9cm]{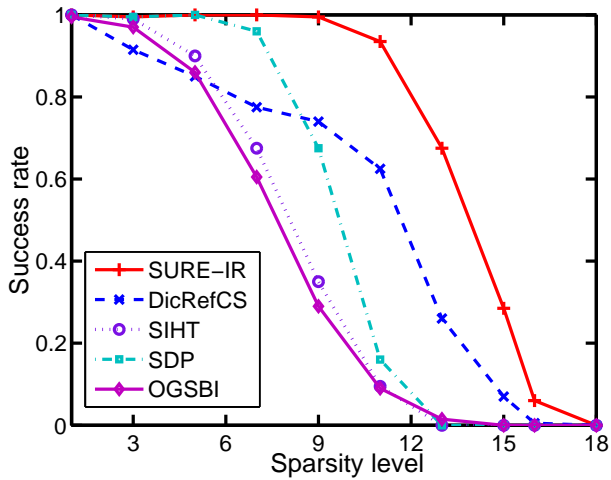}}
  \caption{RSNRs and success rates of respective algorithms vs. $K$, $T=64$, $M=30$, and
$\text{PSNR}=25\text{dB}$.}
   \label{fig2}
\end{figure*}

\begin{figure*}[!t]
 \centering
\subfigure[RSNRs vs.
PSNR.]{\includegraphics[width=9cm]{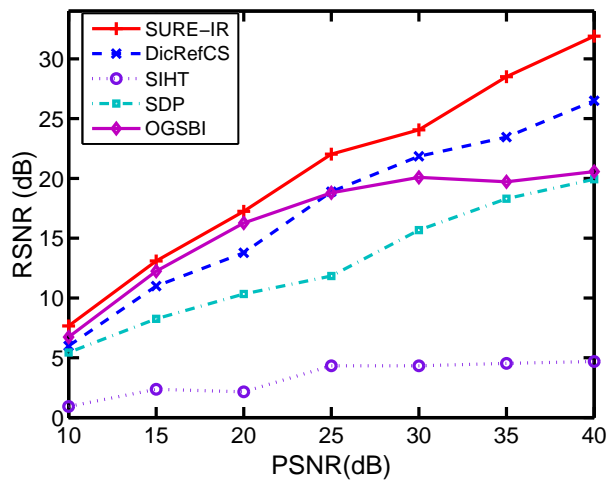}}
 \hfil
\subfigure[Success rates vs.
PSNR.]{\includegraphics[width=9cm]{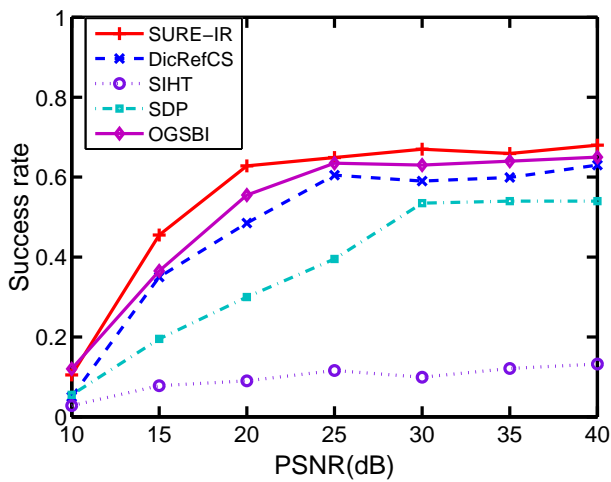}}
  \caption{RSNRs and success rates of respective algorithms vs. PSNR, $T=64$, $M=10$, and
$K=3$.}
   \label{fig3}
\end{figure*}

In the following, we examine the behavior of the respective
algorithms under different scenarios. In Fig. \ref{fig1}, we plot
the average RSNRs and success rates of respective algorithms as a
function of the number of measurements $m$, where we set $T=64$,
$K=3$, $\text{PSNR}=25\text{dB}$. Results are averaged over $10^3$
independent runs, with $\{\alpha_k\}$, $\{\omega_k\}$ and the
sampling indices (used to obtain $\boldsymbol{y}$ from
$\boldsymbol{y}_T$) randomly generated for each run. We see that
the proposed method is superior to all other four methods in terms
of both the RSNR and success rate. In particular, it is worth
mentioning that the proposed method outperforms the SDP method
which is guaranteed to find the global solution. This is probably
because the log-sum penalty functional adopted by our algorithm is
more sparse-encouraging than the atomic norm that is considered as
the continuous analog to the $\ell_1$ norm for discrete signals.
We also observe that the SIHT method yields poor performance for
small $M$, mainly because the embedded subspace-based method
(MUSIC or ESPRIT) for line spectral estimation barely works with a
small number of data samples. The performance of the SIHT method,
however, improves dramatically as $M$ increases. Moreover, we see
that the OGSBI method, though using a very fine grid, still
achieves performance inferior to the proposed SURE-IR and DicRefCS
methods. In Fig. \ref{fig2}, we depict the RSNRs and success rates
of respective algorithms vs. the number of complex sinusoids, $K$,
where we set $T=64$, $M=30$, and $\text{PSNR}=25\text{dB}$. It can
be observed that our proposed SURE-IR algorithm outperforms other
methods by a big margin for a moderately large number of complex
sinusoids $K$. For example, when $K=10$, a gain of over
$10\text{dB}$ in RSNR can be achieved by our algorithm as compared
with the DicRefCS and the SDP methods. This advantage makes our
algorithm the most attractive for scenarios consisting of a
moderate or large number of sinusoid components. The recovery
performance of respective algorithms under different peak
signal-to-noise ratios (PSNRs) is plotted in Fig. \ref{fig3},
where we choose $M=10$, $T=64$, and $K=3$. We see that our
proposed SURE-IR method presents uniform superiority over other
methods under different PSNRs.


\begin{figure*}[!t]
 \centering
\subfigure[RSNRs vs.
$\mu$.]{\includegraphics[width=9cm]{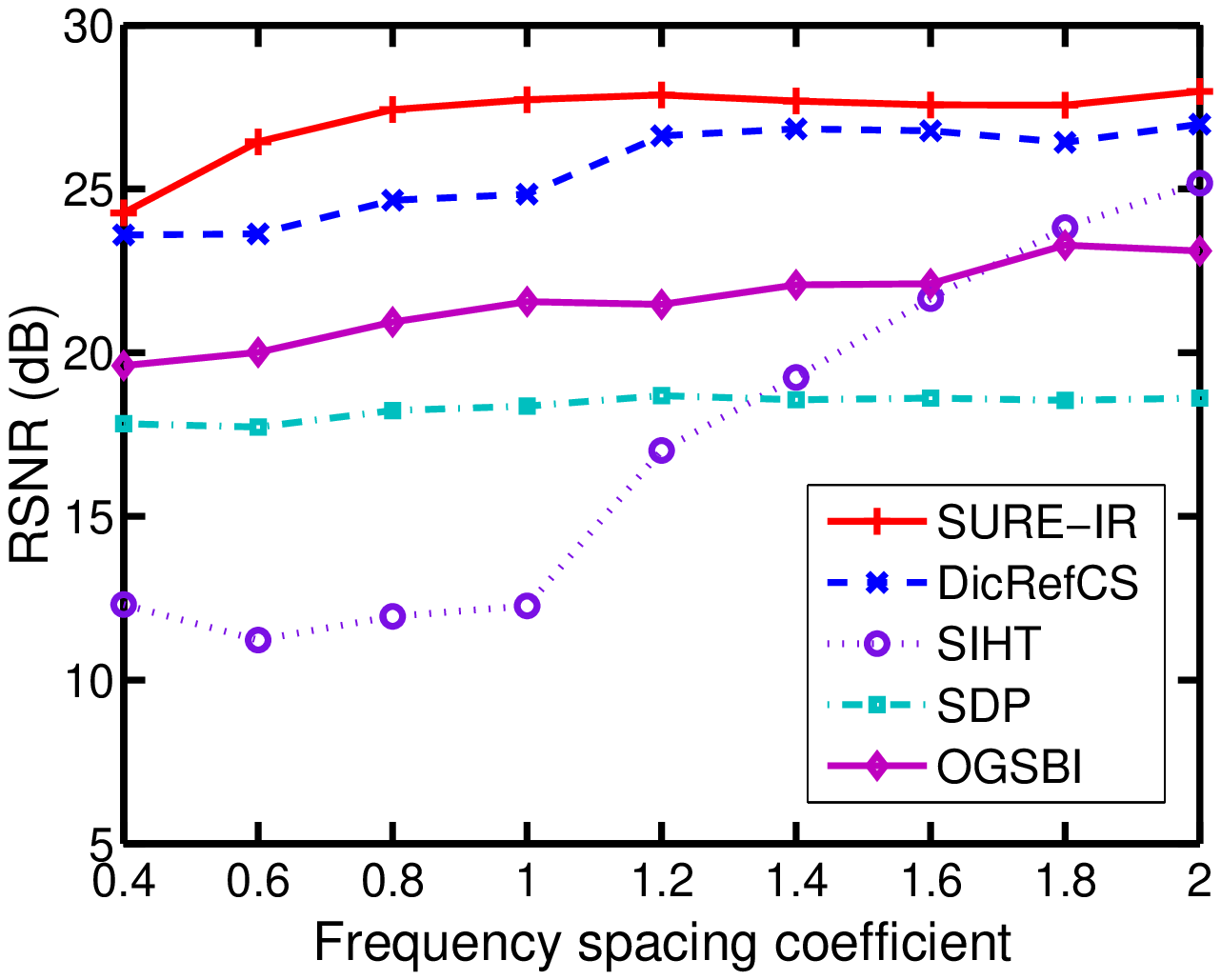}}
 \hfil
\subfigure[Success rates vs.
$\mu$.]{\includegraphics[width=9cm]{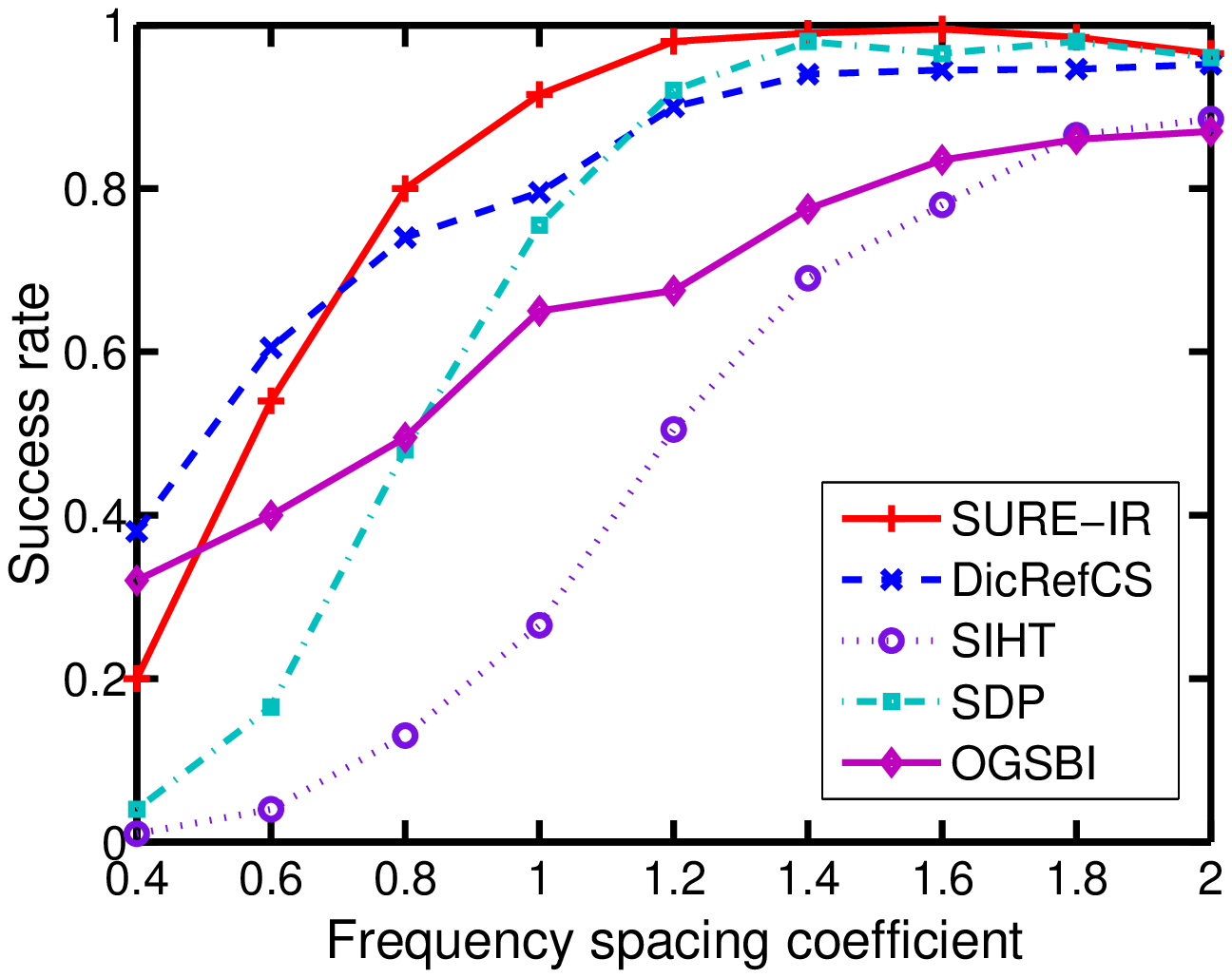}}
  \caption{RSNRS and success rates of respective algorithms vs. the frequency spacing coefficient $\mu$, $T=64$, $M=20$, and
$\text{PSNR}=15\text{dB}$.}
   \label{fig4}
\end{figure*}

We now examine the ability of respective algorithms in resolving
closely-spaced frequency components. The signal $\boldsymbol{y}$
is assumed a mixture of two complex sinusoids with the frequency
spacing $d_f\triangleq\frac{1}{2\pi}(\omega_1-\omega_2)=\mu/T$,
where $\mu$ is the frequency spacing coefficient ranging from
$0.4$ to $2$. Fig. \ref{fig4} shows RSNRs and success rates of
respective algorithms vs. the frequency spacing coefficient $\mu$,
where we set $T=64$, $M=20$, and $\text{PSNR}=15\text{dB}$.
Results are averaged over $10^3$ independent runs, with one of the
two frequencies (the other frequency is determined by the
frequency spacing) and the set of sampling indices randomly
generated for each run. We observe that when the two frequency
components are very close to each other, e.g. $\mu=0.6$, the SDP
and the SIHT can hardly identify the true frequency parameters,
whereas the SURE-IR and the DicRefCS are still capable of
resolving these two closely-spaced components with decent success
rates. Although the DicRefCS method slightly outperforms (in terms
of the success rate) the SURE-IR method in the very small
frequency spacing regime, it is quickly surpassed by the SURE-IR
method as the frequency spacing increases.



\begin{figure}[!t]
\centering
\includegraphics[width=9cm]{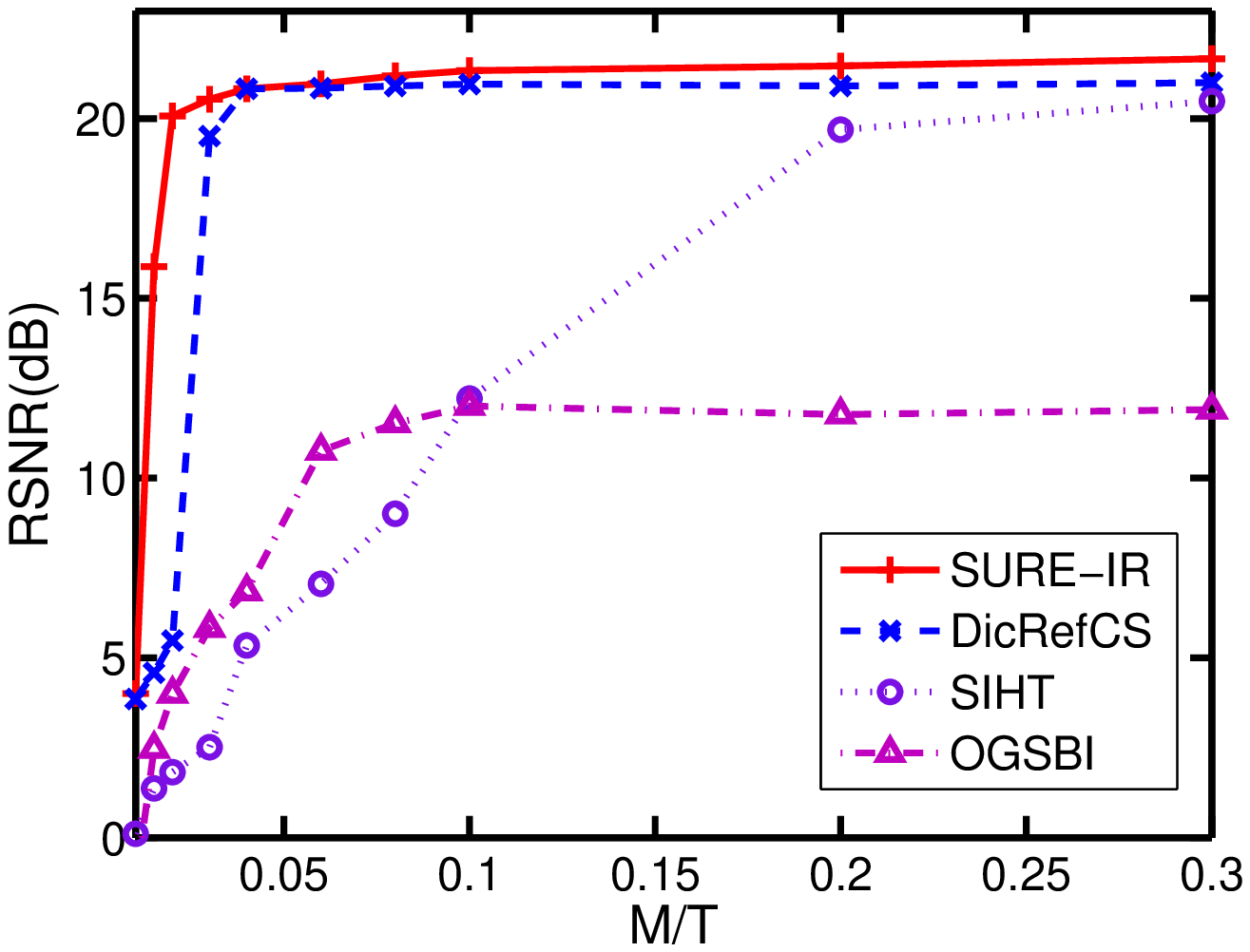}
\caption{RSNRs of respective algorithms vs. the ratio $M/T$.}
\label{fig5}
\end{figure}

\begin{figure*}[!t]
 \centering
\begin{tabular}{ccc}
\hspace*{-3ex}
\includegraphics[width=5cm]{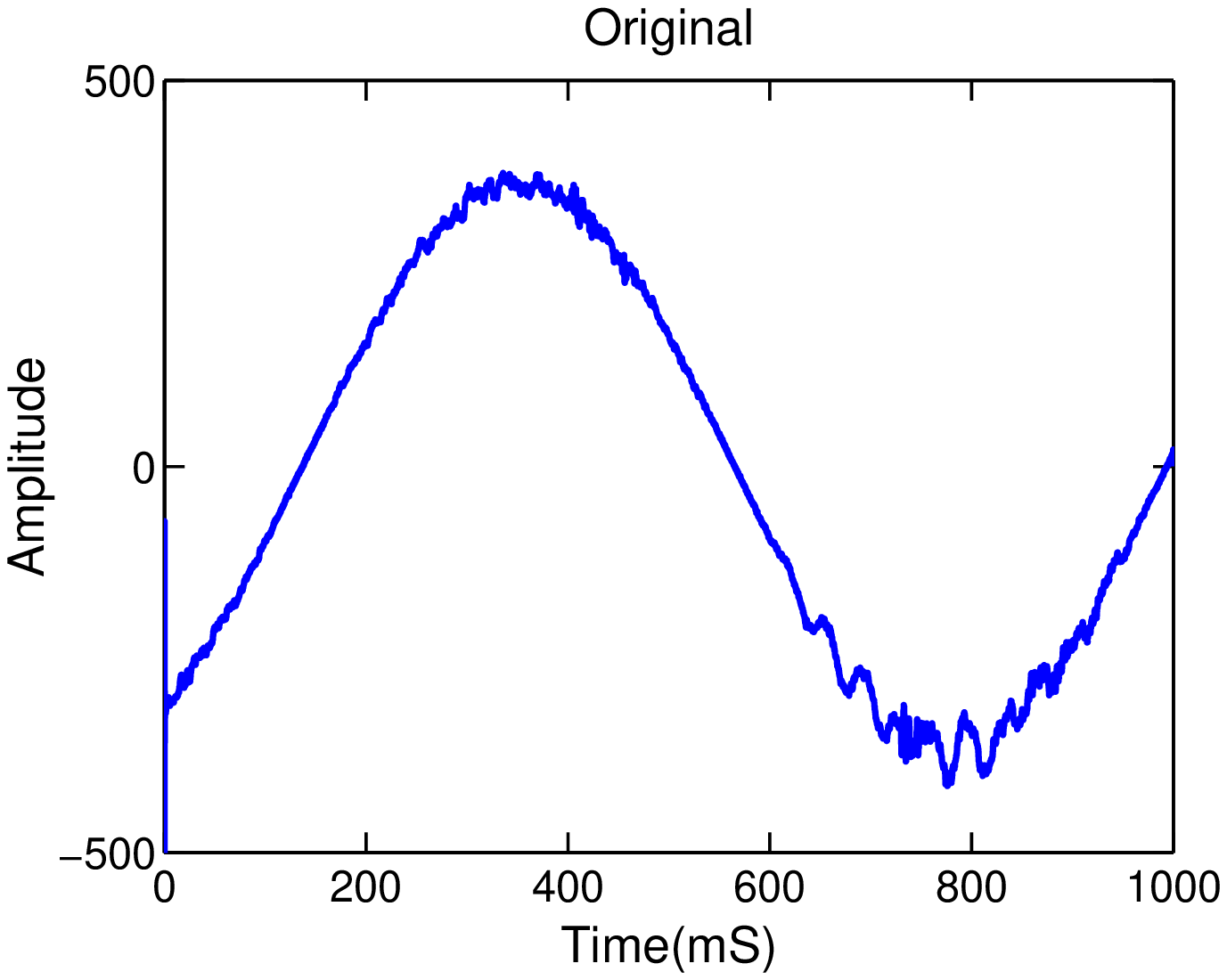}&
\hspace*{-6ex}
\includegraphics[width=5cm]{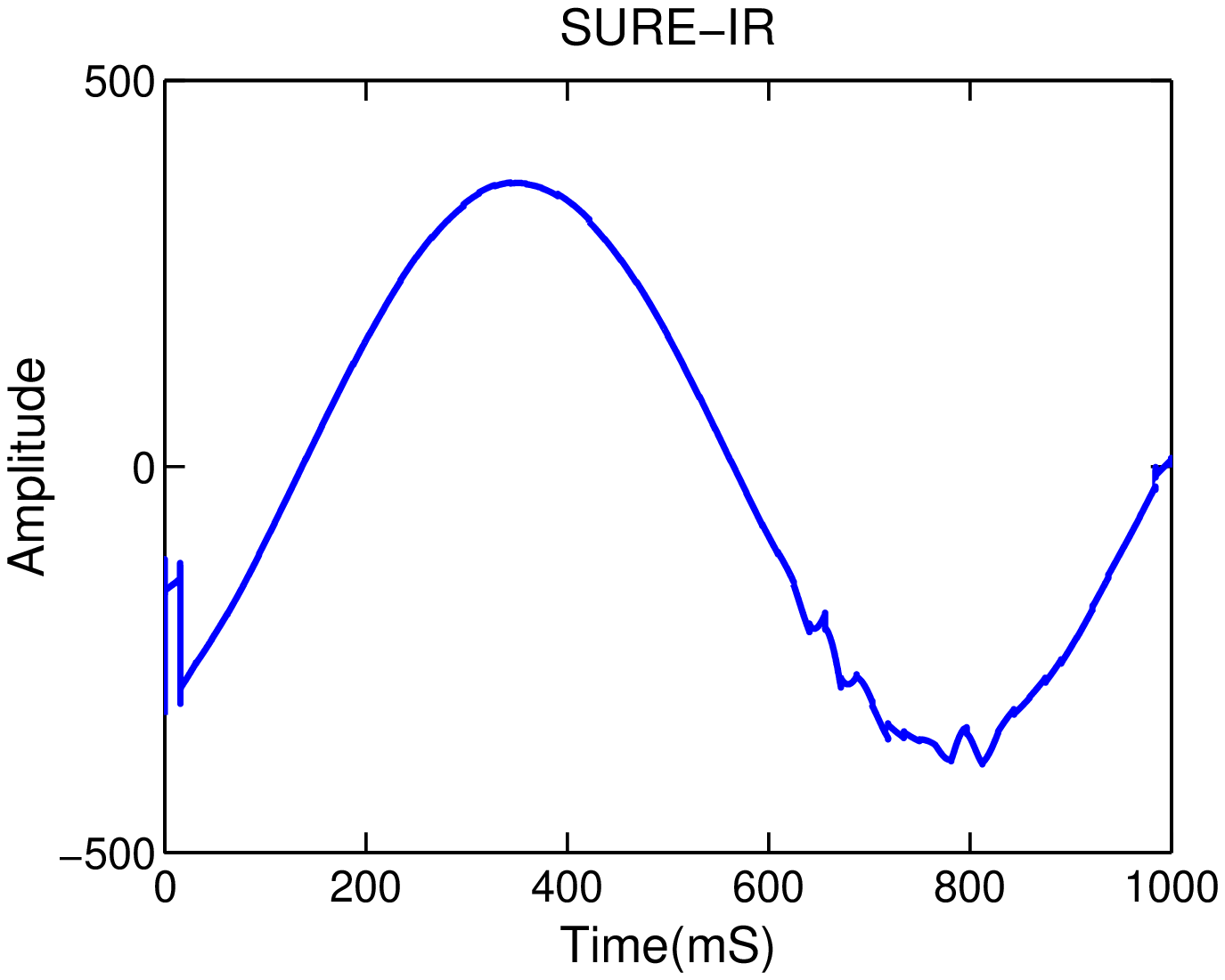}
& \hspace*{-6ex}
\includegraphics[width=5cm]{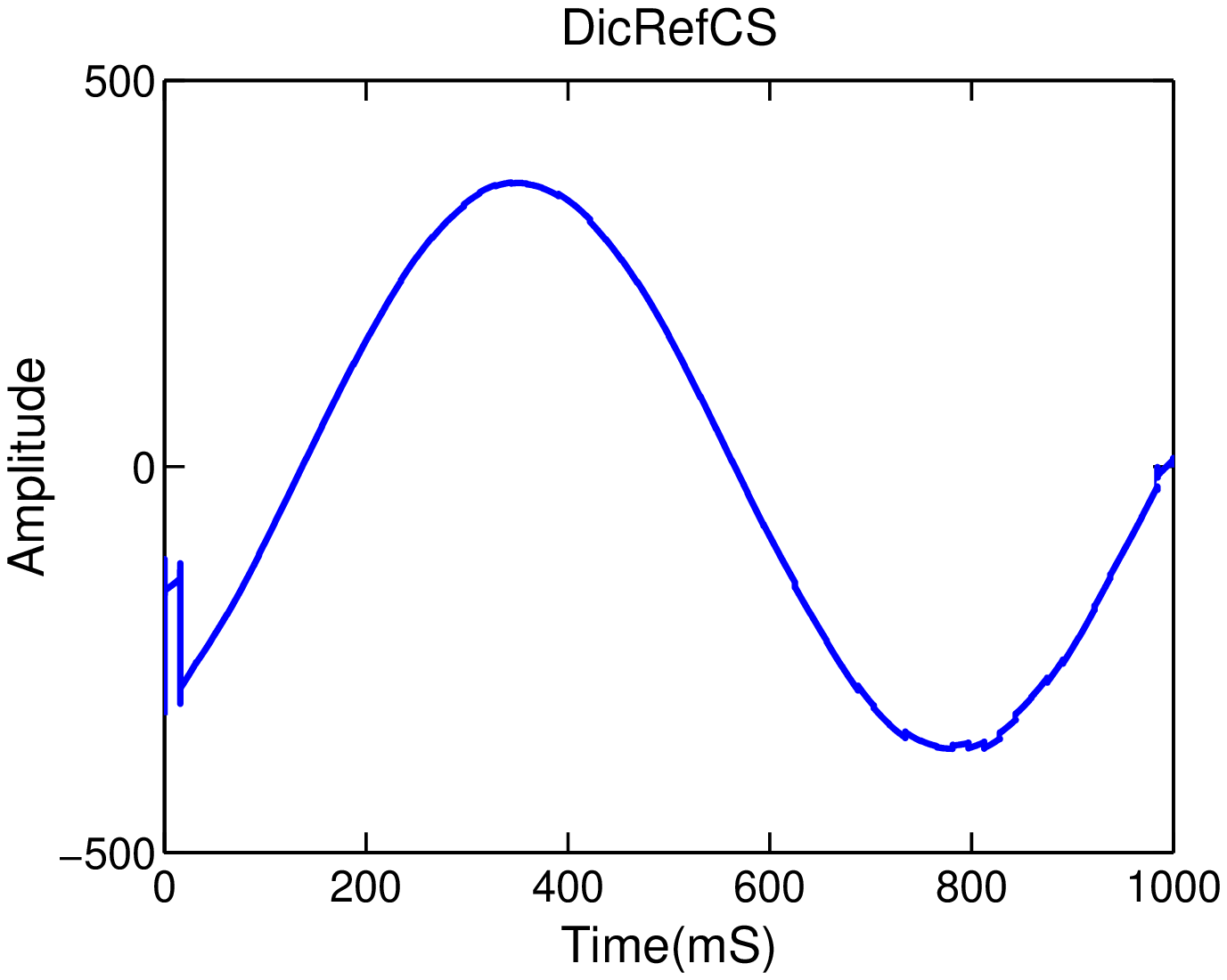}
\\
\hspace*{-3ex}\includegraphics[width=5cm]{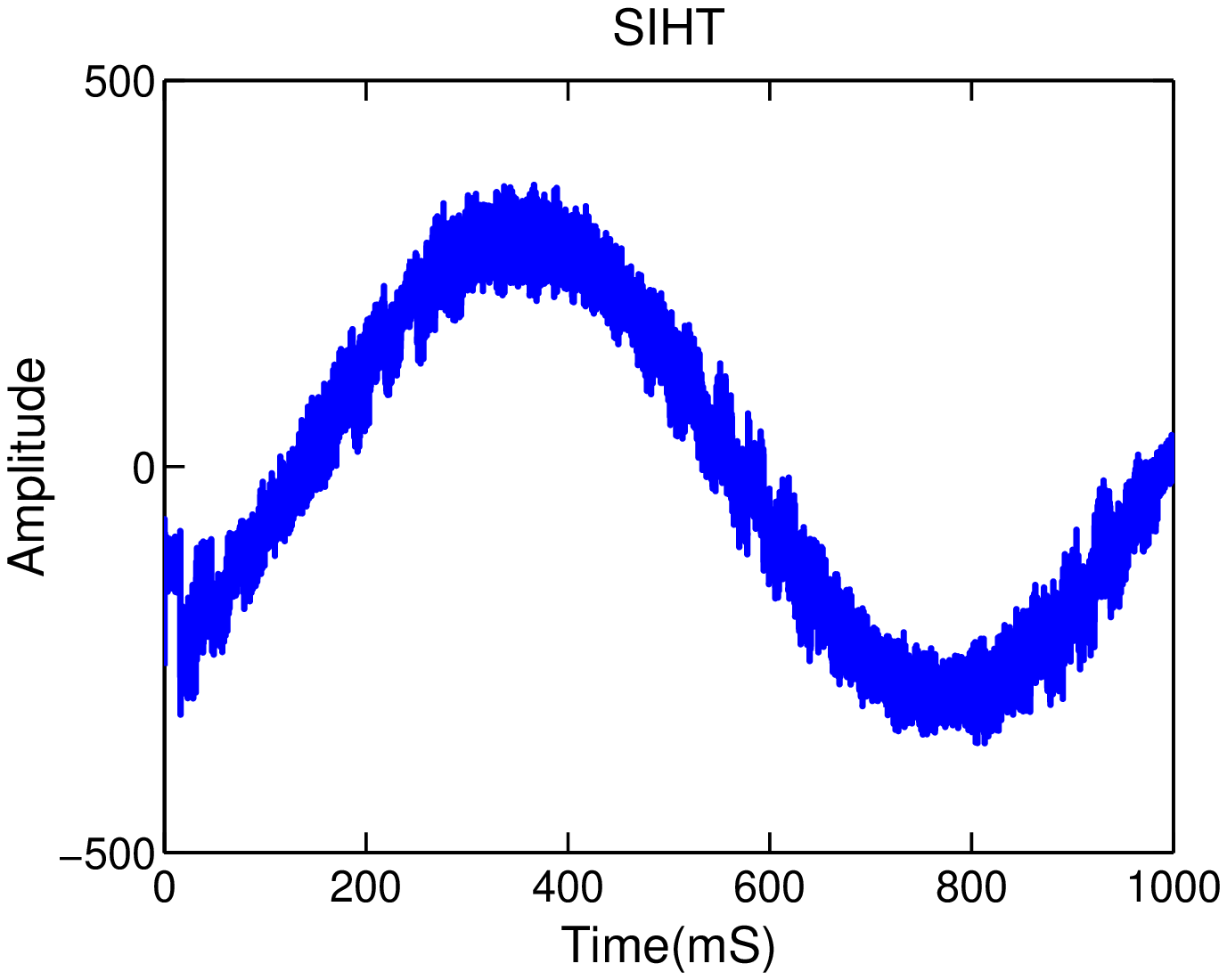} & \hspace*{-6ex}
\includegraphics[width=5cm]{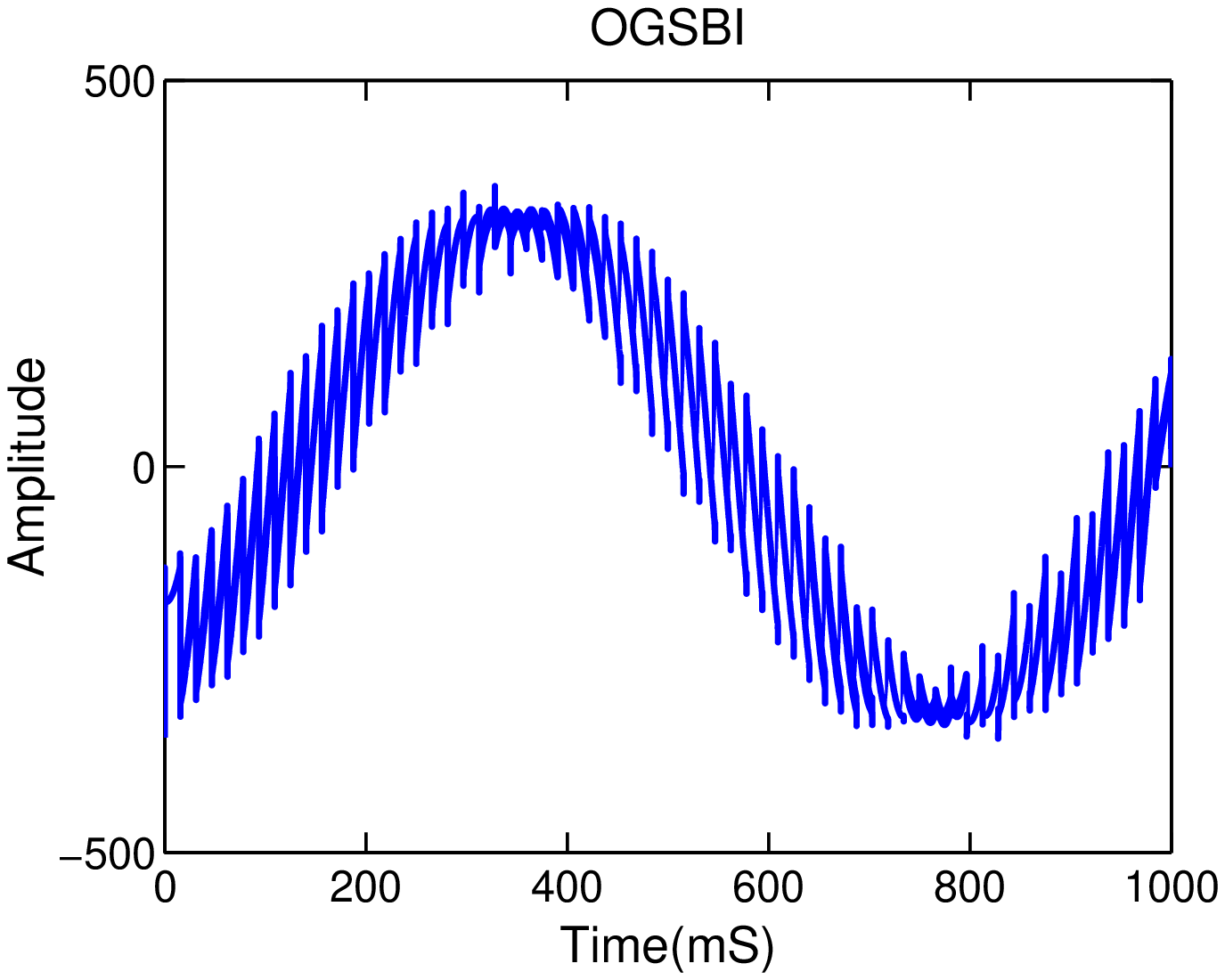}
\end{tabular}
  \caption{The true message and the
messages reconstructed by respective algorithms, $M=100$.}
   \label{fig6}
\end{figure*}

\begin{figure*}[!t]
 \centering
\begin{tabular}{ccc}
\hspace*{-3ex}
\includegraphics[width=5cm]{original}&
\hspace*{-6ex}
\includegraphics[width=5cm]{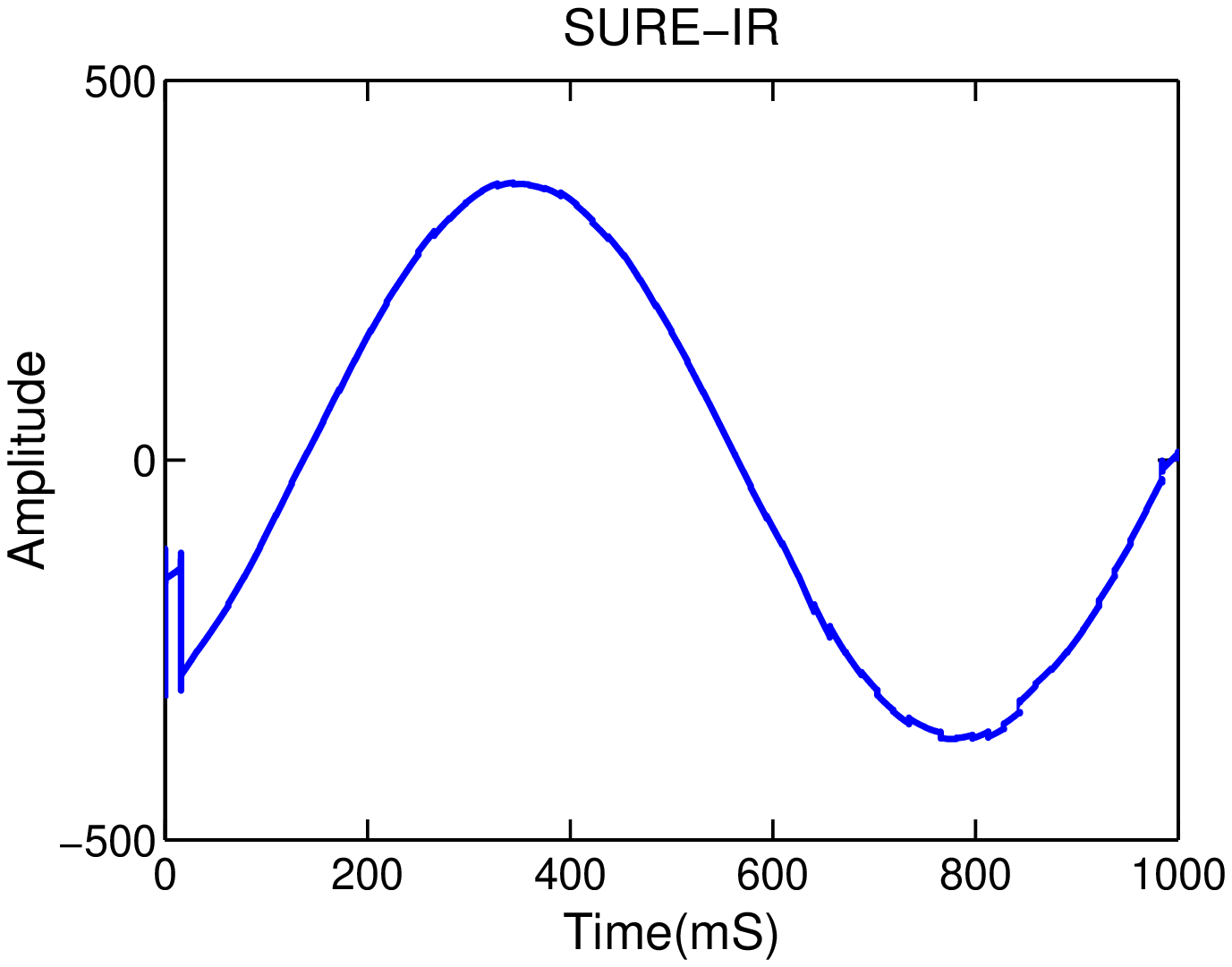}
& \hspace*{-6ex}
\includegraphics[width=5cm]{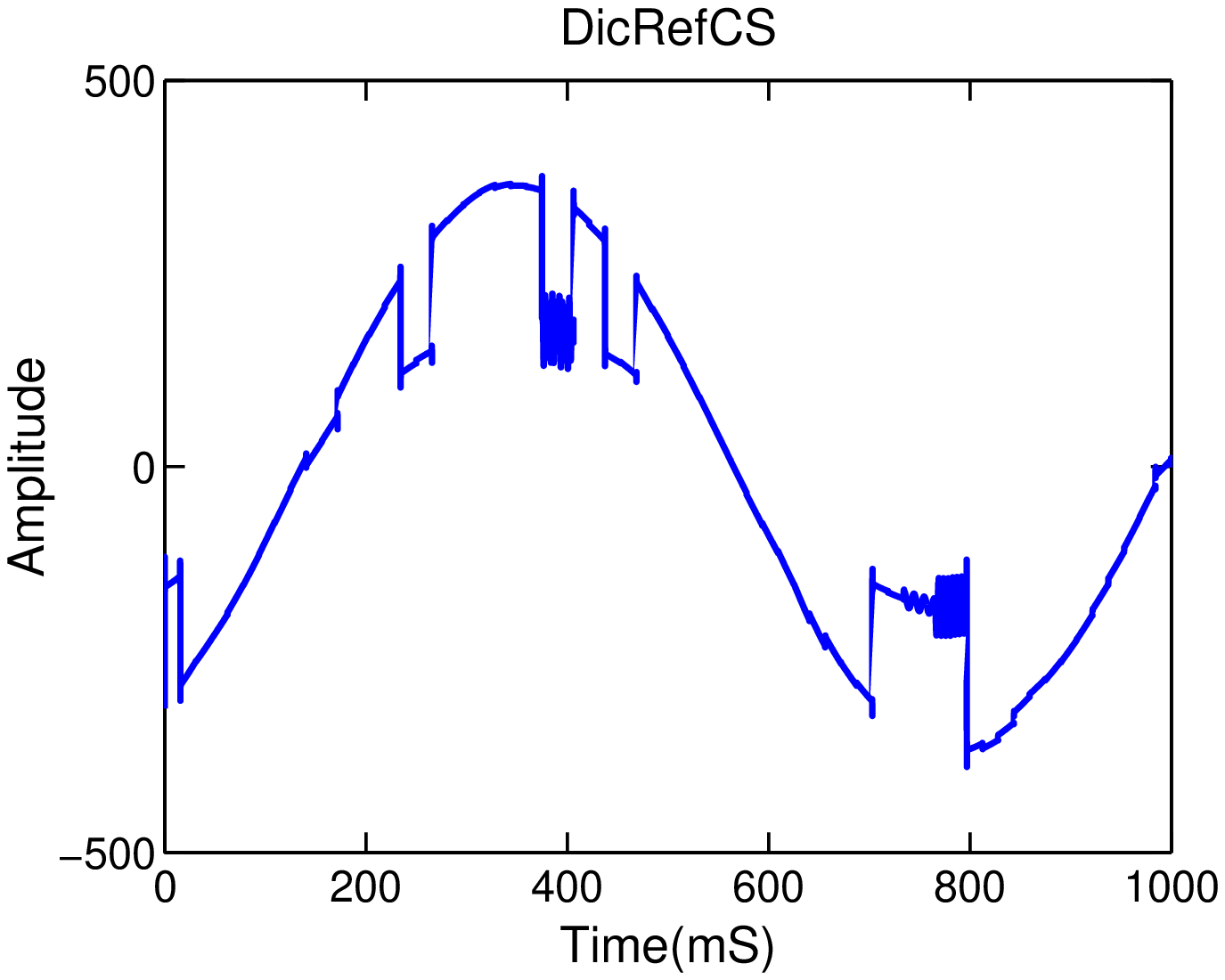}
\\
\hspace*{-3ex}\includegraphics[width=5cm]{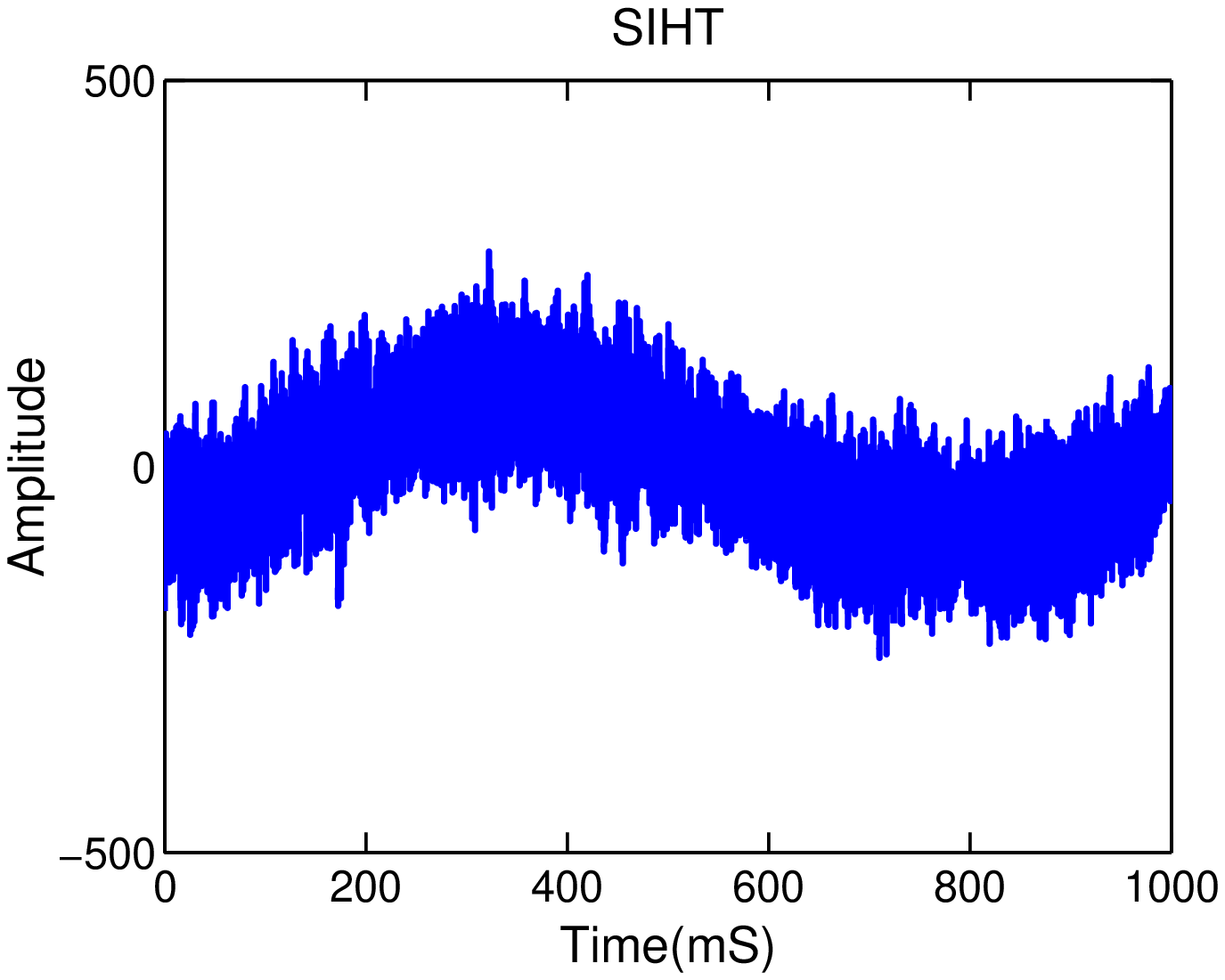} & \hspace*{-6ex}
\includegraphics[width=5cm]{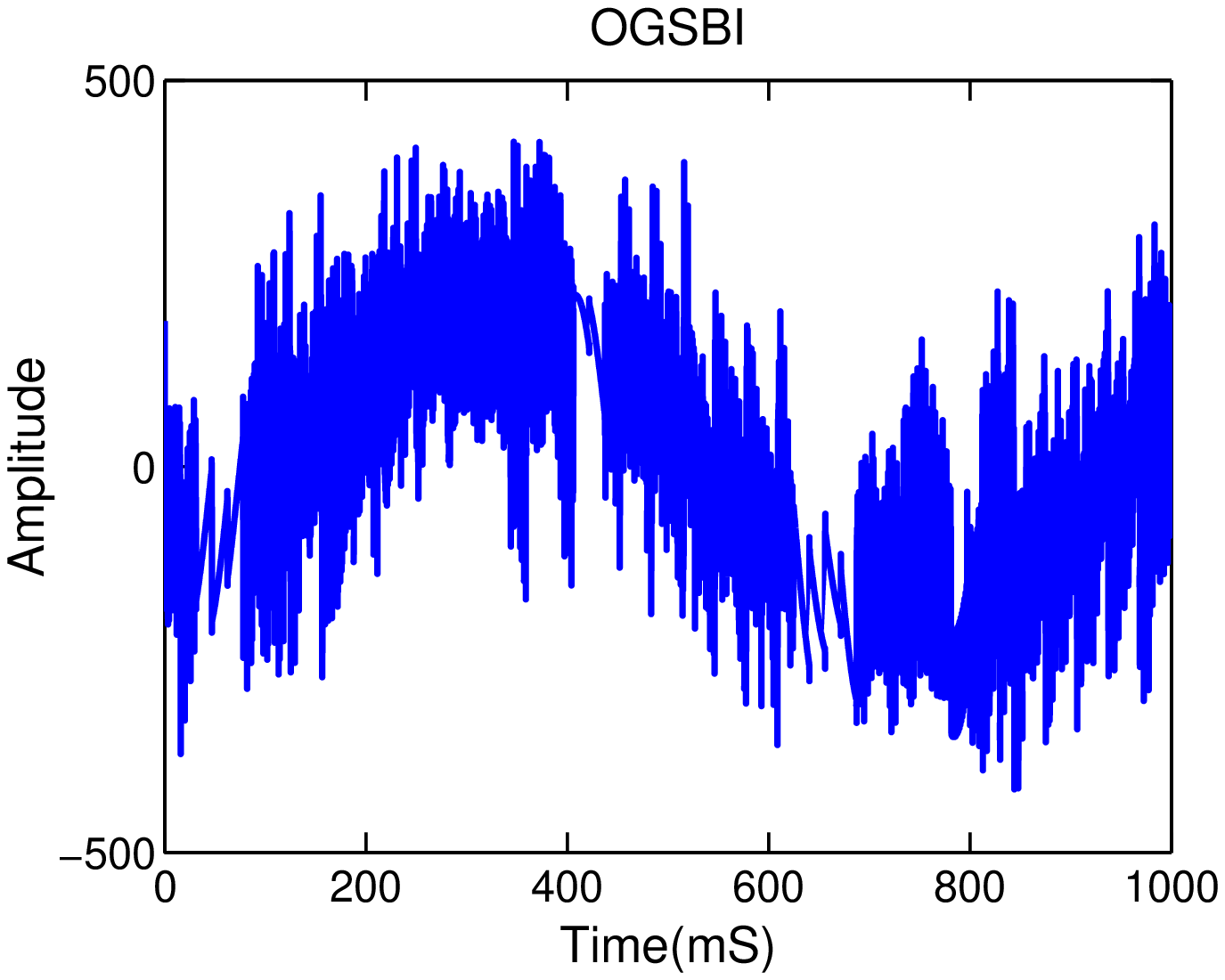}
\end{tabular}
  \caption{The true message and the
messages reconstructed by respective algorithms, $M=20$.}
   \label{fig7}
\end{figure*}

Our last experiment tests the recovery performance of respective
algorithms using a real-world amplitude modulated (AM) signal
\cite{TroppLaska10,DuarteBaraniuk13} that encodes the message
appearing in the top left corner of Fig. \ref{fig6}. The signal
was transmitted from a communication device using carrier
frequency $\omega_c=8.2$kHz, and the received signal was sampled
by an analog-digital converter (ADC) at a rate of $32$kHz. The
sampled signal has a total number of $32768$ samples. For the sake
of computational efficiency, in our experiment, the AM signal is
divided into a number of short-time segments, each consisting of
$T=1024$ data samples. For each segment, we randomly select $M$
data samples, based on which we use respective algorithms to
recover the whole segment. After all segments are reconstructed,
we perform AM demodulation on the recovered signal to reconstruct
the original message. The RSNR is then computed using the
reconstructed message and the true message. Fig. \ref{fig5} plots
the RSNRs of respective algorithms vs. the ratio $M/T$ (the SDP
method was not included in this experiment due to its prohibitive
computational complexity when the signal dimension is large). We
see that our proposed SURE-IR method offers the best performance
and presents a significant performance advantage over the other
algorithms for a small ratio $M/T$, where data acquisition is more
beneficial due to high compression rates. In particular, when
$M/T=0.02$, all the other three methods (DicRefCS, SIHT and OGSBI)
fail to provide an accurate reconstruction, while our proposed
algorithm still renders a decent recovery accuracy. Figs.
\ref{fig6} and \ref{fig7} show the true message and the messages
recovered by respective algorithms, where $M$ is set to $20$ and
$100$, respectively. It can be seen that our proposed algorithm
can obtain a fairly accurate reconstruction of the original signal
even with as few as $M=20$ measurements, whereas the messages
reconstructed by the SIHT and the OGSBI methods are highly
smeared/distorted.


\section{Conclusions} \label{sec:conclusion}
This paper studied the super-resolution compressed sensing problem
where the sparsifying dictionary is characterized by a set of
unknown parameters in a continuous domain. Such a problem arises
in many practical applications such as direction-of-arrival
estimation and line spectral estimation. By resorting to the
majorization-minimization approach, we developed a generalized
iterative reweighted $\ell_2$ algorithm for joint dictionary
parameter learning and sparse signal recovery. The proposed
algorithm iteratively decreases a surrogate function majorizing a
given objective function, leading to a gradual and interweaved
iterative process to refine the unknown parameters and the sparse
signal. Simulation results show that our proposed algorithm
effectively overcomes the grid mismatch problem and achieves a
super-resolution accuracy in resolving the unknown frequency
parameters. The proposed algorithm also demonstrates superiority
over several existing super-resolution compressed sensing methods
in resolving the unknown parameters and reconstructing the
original signal.

\useRomanappendicesfalse
\appendices

\section{Derivative of $f(\boldsymbol{\theta})$ W.R.T. $\boldsymbol{\theta}$} \label{appA}
Define
\begin{align}
\boldsymbol{X} \triangleq \boldsymbol{A}(\boldsymbol{\theta})
\big(\boldsymbol{A}^H(\boldsymbol{\theta})\boldsymbol{A}(\boldsymbol{\theta})
+ \lambda\boldsymbol{D}^{(t)}\big)^{-1}
\boldsymbol{A}^H(\boldsymbol{\theta}) \nonumber
\end{align}
Using the chain rule, the first derivative of
$f(\boldsymbol{\theta})$ with respect to $\theta_i$, $\forall i$
can be computed as
\begin{align}
\frac{\partial f(\boldsymbol{\theta})}{\partial \theta_i} =
\text{tr}\bigg\{ \bigg(\frac{\partial
f(\boldsymbol{\theta})}{\partial \boldsymbol{X}}\bigg)^T
\frac{\partial \boldsymbol{X}}{\partial \theta_i} \bigg\} +
\text{tr}\bigg\{ \bigg(\frac{\partial
f(\boldsymbol{\theta})}{\partial \boldsymbol{X}^\ast}\bigg)^T
\frac{\partial \boldsymbol{X}^\ast}{\partial \theta_i} \bigg\}
\nonumber
\end{align}
where $\boldsymbol{X}^\ast$ donates the conjugate of the complex
matrix $\boldsymbol{X}$, and
\begin{align}
\frac{\partial f(\boldsymbol{\theta})}{\partial \boldsymbol{X}} &=
\frac{\partial}{\partial \boldsymbol{X}}
\text{tr}\{-\boldsymbol{y}\boldsymbol{y}^H \boldsymbol{X}\}
=-(\boldsymbol{y}\boldsymbol{y}^H)^T \nonumber \\
\frac{\partial f(\boldsymbol{\theta})}{\partial
\boldsymbol{X}^\ast} &= \frac{\partial}{\partial
\boldsymbol{X}^\ast} \text{tr}\{ -\boldsymbol{y}\boldsymbol{y}^H
\boldsymbol{X}\}=\boldsymbol{0} \nonumber
\end{align}

\begin{align}
\frac{\partial \boldsymbol{X}}{\partial \theta_i} &=
\frac{\partial}{\partial \theta_i}
\bigg(\boldsymbol{A}(\boldsymbol{\theta})
\big(\boldsymbol{A}^H(\boldsymbol{\theta})\boldsymbol{A}(\boldsymbol{\theta})
 + \lambda^{-1}\boldsymbol{D}^{(t)}\big)^{-1} \boldsymbol{A}^H(\boldsymbol{\theta}) \bigg) \nonumber \\
&= \frac{\partial \boldsymbol{A}(\boldsymbol{\theta})}{\partial
\theta_i}
\big(\boldsymbol{A}^H(\boldsymbol{\theta})\boldsymbol{A}(\boldsymbol{\theta})
+ \lambda^{-1}\boldsymbol{D}^{(t)}\big)^{-1}
\boldsymbol{A}^H(\boldsymbol{\theta})
\nonumber \\
&\quad +\boldsymbol{A}(\boldsymbol{\theta})
\big(\boldsymbol{A}^H(\boldsymbol{\theta})\boldsymbol{A}(\boldsymbol{\theta})
+ \lambda^{-1}\boldsymbol{D}^{(t)}\big)^{-1}
\frac{\partial\boldsymbol{A}^H(\boldsymbol{\theta})}{\partial\theta_i}
\nonumber \\
&\quad
+\boldsymbol{A}(\boldsymbol{\theta})\big(\boldsymbol{A}^H(\boldsymbol{\theta})\boldsymbol{A}(\boldsymbol{\theta})
+\lambda^{-1}\boldsymbol{D}^{(t)}\big)^{-1}
\bigg(\frac{\partial\boldsymbol{A}^H(\boldsymbol{\theta})}{\partial\theta_i}\boldsymbol{A}(\boldsymbol{\theta})
\nonumber \\
&\quad
+\boldsymbol{A}^H(\boldsymbol{\theta})\frac{\partial\boldsymbol{A}(\boldsymbol{\theta})}{\partial\theta_i}\bigg)
\big(\boldsymbol{A}^H(\boldsymbol{\theta})\boldsymbol{A}(\boldsymbol{\theta})+\lambda^{-1}\boldsymbol{D}^{(t)}\big)^{-1}
\boldsymbol{A}^H(\boldsymbol{\theta}) \nonumber
\end{align}


\end{document}